# Nanoconfined Grain Boundaries Increase Conductivity of Polycrystalline Molecular Crystals

*Shujit Chandra Paul[1], William A. Goddard III,[2] Michael J. Zdilla,*[1] Prabhat Prakash*[2] and Stephanie L. Wunder*[1]*

[1]Department of Chemistry, Temple University, Philadelphia, PA 19122, United States

[2]Materials and Process Simulation Center (MSC), California Institute of Technology, Pasadena, California, 91125, United States

*Corresponding Authors Email: michael.zdilla@temple.edu; pprakash@caltech.edu; slwunder@temple.edu

## ABSTRACT

Soft-solid molecular crystals consist of crystalline grains and fluid grain boundaries (GB) that enhance grain binding and transport of $Li^+$ ions between the grains. The total ionic conductivity consists of ion migration in both the grains and GBs. To unravel these contributions in adiponitrile (Adpn):$LiPF_6$ molecular crystals, the GB volume fraction was varied by changing the size of the crystals and the Adpn:$LiPF_6$ molar ratio. Molecular dynamic (MD) simulations indicate that ion motion was sub-diffusive in the grains and "well-diffusive" in the GBs, with GBs characterized as disordered nano-confined regions of higher charge carrier concentration (~1M) than in saturated Adpn:$LiPF_6$ solutions (0.04M), and $Li^+$ ions predominantly solvated by cyano groups with few contact ion pairs. The diffusivity in the GBs is at least an order of magnitude higher than in the crystalline grains. The emergent picture is the grains as a reservoir of ions that migrate to the faster-conducting GBs.

Solid electrolytes provide safety advantages over liquid electrolytes and have thus been investigated for use in Li metal (LMBs), lithium-ion (LIBs) and next generation batteries. The solid electrolytes investigated have been inorganic lithium-ion conductive ceramics (LICCs)[1,2], polymer or polymer gel electrolytes[3], or composite electrolytes with inorganic and organic components. LICCs can have conductivities greater than liquid electrolytes, but are brittle, with resistive grain boundaries[4], while polymer electrolytes are flexible and malleable but have poor ionic conductivity, the latter of which can be enhanced by addition of liquids to form gels. Composite electrolytes were proposed to combine advantages of the LICCs and polymers[5]. However, there appears to be a barrier to $Li^+$ ion migration between the inorganic and organic phase, with the resistance at the interface limiting the total ionic conductivity.

Enhanced ionic conductivity in soft matter has been observed in materials that exhibit some structural order. Both liquid crystals(discs), which have no long range order (with no or very broad Bragg peaks) between the crystal and melt phases, and plastic crystals (spheres, with low melt entropy, and which exhibit strong long range order), can contain mobile ions with enhanced ionic transport properties in these phases.[6] Liquid-crystalline (LC) columnar assemblies[7, 8] based on ionic liquids[9], doped ionic liquids[10], liquid crystalline salts[11], ion doped plastic crystals[12-21] , and materials with both liquid crystalline and plastic crystal phases[10], form conductive phases. Enhanced ionic conductivity is observed in membranes containing oriented supramolecular transport channels[22] and in polymer hydrogels.[23] Ion channels form in molecular crystalline soft-solid electrolytes with stoichiometric ratios of organics/salts such as sulfones[24], diamines[25], diamides[26], ethers[25, 27], glymes[28-32] and dinitriles.[33] Trilithium crystalline LiTFSI/organic complexes[31], self-assembled phthalocyanines and related compounds[34], and crown ethers[35-37] also form ion conduction paths in stacked channel structures.

In the case of soft-solid molecular crystals, there are several factors that influence ionic conductivity. These include: (i) the occurrence of $Li^+$ ion channels with no contact ion pairs between the anions and cations to provide directional motion[38]; (ii) more than one channel (2D or 3D migration) preferable to one channel (1D motion); (iii) short $Li^+\cdots Li^+$ hopping distances[39]; and (iv) weak ligand··· $Li^+$ interactions with soft ligands (e.g. C≡N) preferable to hard (e.g. ether oxygens)[40]. In addition to these considerations, we show here that grain boundaries play an important role. Grain boundaries in molecular crystals are fluid-like and promote $Li^+$ ion diffusion and adhesion between the grains.[38, 41]

Here we investigate the conductive interphases between grains by comparing off-stoichiometric $(Adpn)_{2.3}LiPF_6$ molecular crystals containing excess Adpn, to stoichiometric $(Adpn)_2LiPF_6$ molecular crystals of different sizes **(Figure 1)**. It is important to note that the 15% excess Adpn does not macrophase separate but rather resides in the nano-confined region between the grains. PXRD data **(Figure S1 in SI)** are identical for all the samples and agree with PXRD generated from the single crystal XRD of $(Adpn)_2LiPF_6$ with no evidence of Adpn, which at T = 100K is amorphous. Crystal sizes, and thus the amount of grain boundary contribution, were varied: Small crystals were obtained by melt crystallization or evaporation from THF and variable size crystals by slower evaporation from AN (**Figure 1**, **Figure S2, S3**). Thermogravimetric analysis (**TGA**) data (**Figure S4, Table S1**) shows excellent agreement between the experimental and expected weight loss values for the added Adpn. Differential scanning calorimetry data (**Figure 2A, Table 1**) show that the width of the melt peaks for the melt crystallized $(Adpn)_2LiPF_6$, $(Adpn)_{2.3}LiPF_6$ and THF solution crystallized $(Adpn)_2LiPF_6$ is narrower, with melt temperatures, $T_m \approx 178\ ^0C$ and melt enthalpies, $\Delta H_m \approx 178$ J/g that are higher by 5ºC and ~ 5 J/g than all the AN solution crystallized samples with $T_m \approx 173\ ^0C$, $\Delta H_m \approx 172$ J/g. These differences indicate more perfect crystal formation in the former case. For the $(Adpn)_{2.3}LiPF_6$ crystals, additional small melting peaks are observed at $T_m$ (Adpn) ~ -1 to -4.3 ºC with small crystallization endotherms (15.8 J/g).

Higher $T_m$ values **(Figure 2A)** can indicate either larger or more perfect crystals. However, here the melt crystallized small crystals (with or without excess Adpn) have higher values of $T_m$ than the similarly small-sized AN solution crystallized sample (with or without excess Adpn). Both the large and small AN solution crystallized $(Adpn)_2LiPF_6$ and $(Adpn)_{2.3}LiPF_6$ have the same lower $T_m$. Therefore, the crystal size does not account for the 5 $^0C$ decrease in $T_m$ values for all the samples crystallized from AN. In addition to the higher $T_m$ for the melt crystallized samples, the supercooling temperature, i.e., the difference between the melt and crystallization temperature ($\Delta T = T_m - T_c$) is less for the small melt crystallized $(Adpn)_2LiPF_6$ ($\Delta T = 8\ ^0C$) than for the AN crystallized $(Adpn)_2LiPF_6$ small and large samples ($\Delta T = 14\ ^0C$). This suggests that the grains in the melt crystallized samples have fewer defects than those crystallized from AN, making it easier to recrystallize. Defects may occur in the crystals if a mononitrile like AN occupies sites in the crystal lattice that are not completely replaced by the dinitrile Adpn (creating a hole or CIP), or remain in the crystal lattice, either of which can create defect sites that block $Li^+$ ion migration.

To test this hypothesis, the $(Adpn)_2LiPF_6$ crystals grown via evaporation from THF were compared with crystals grown from AN. $^1H$ NMR data of the small (**Figure S5A**) and large (**Figure S5B**) $(Adpn)_2LiPF_6$ molecular crystals in $d_6$-DMSO indicates the presence of residual AN. Dissolution of the $(Adpn)_2LiPF_6$ crystals grown from THF in $d_6$-DMSO showed no THF peaks (**Figure S5C**), i.e. there was no residual THF in the $(Adpn)_2LiPF_6$. These results are consistent with the melting point trends. The high $T_m$ for the more perfect cocrystals grown from THF, without trapped THF, were the same as the $(Adpn)_2LiPF_6$ obtained by melt crystallization (177 $^0$C). Less perfect crystals with defects grown from AN had trapped AN in the grains and $T_m$ values lower by 5 $^0$C. The larger mass and size of the THF ligand may be harder to incorporate into the crystal structure than the smaller AN.

To assess the approximate molarity of $LiPF_6$ in the GBs, melt temperatures of Adpn and $LiPF_6$ solutions were measured (**Figure 2B**) and compared with $T_m$s for the excess Adpn (~ 0 ºC) in the $(Adpn)_{2.3}LiPF_6$ molecular crystals (**Table 1**). In the molecular crystals, Adpn melts at lower temperatures (-1 ºC to -4.3 ºC) compared to neat Adpn (1 ºC to 3 ºC), indicating the presence of dissolved $LiPF_6$. In the $LiPF_6$:Adpn solutions (note: the solubility of $LiPF_6$ in Adpn is very low <0.05 M), the melting points decrease with respect to that of Adpn as the concentration of $LiPF_6$ increases, as expected. However, the melting point of the Adpn in the non-stoichiometric $(Adpn)_{2.3}LiPF_6$ molecular crystal is substantially lower than in the dilute solutions. Since melting point depression is a colligative property, extrapolation of the $T_m$ vs molarity plot, using 0, 0.01, 0.02, 0.03 and 0.04 M, to 1 $^0$C occurred at 0.4 M $LiPF_6$. This result strongly suggests that the excess Adpn is more concentrated in the "nanoconfined" swollen grain boundary regions than in a 0.04 M Adpn solution.

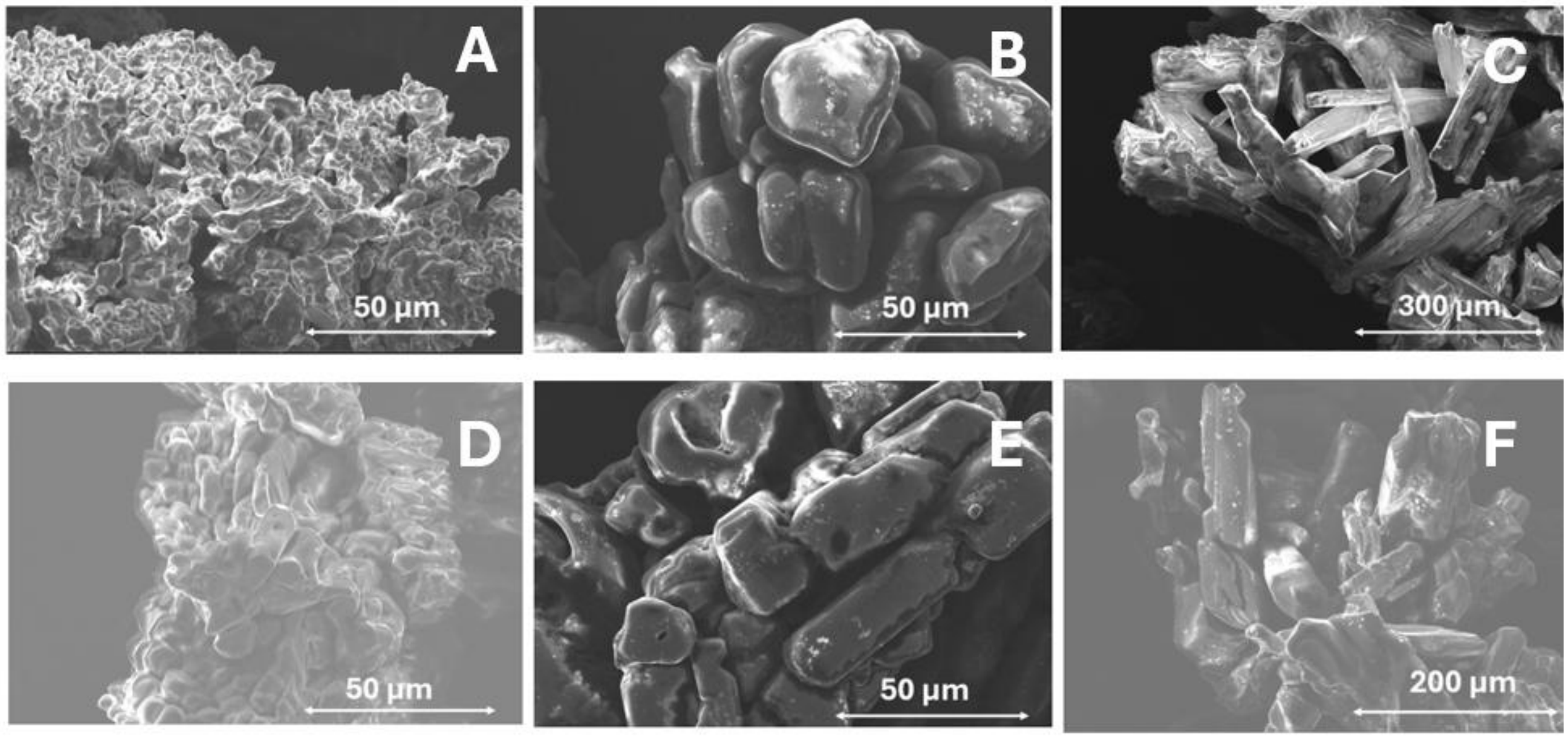

**Figure 1.** SEM Images of $(Adpn)_2LiPF_6$ crystals formed by **(A)** melt crystallization; **(B)** fast crystallization from acetonitrile (AN); **(C)** slow crystallization from AN; and SEM images of $(Adpn)_{2.3}LiPF_6$ crystals formed by **(D)** melt crystallization; **(E)** fast crystallization from AN; and **(F)** slow crystallization from AN.

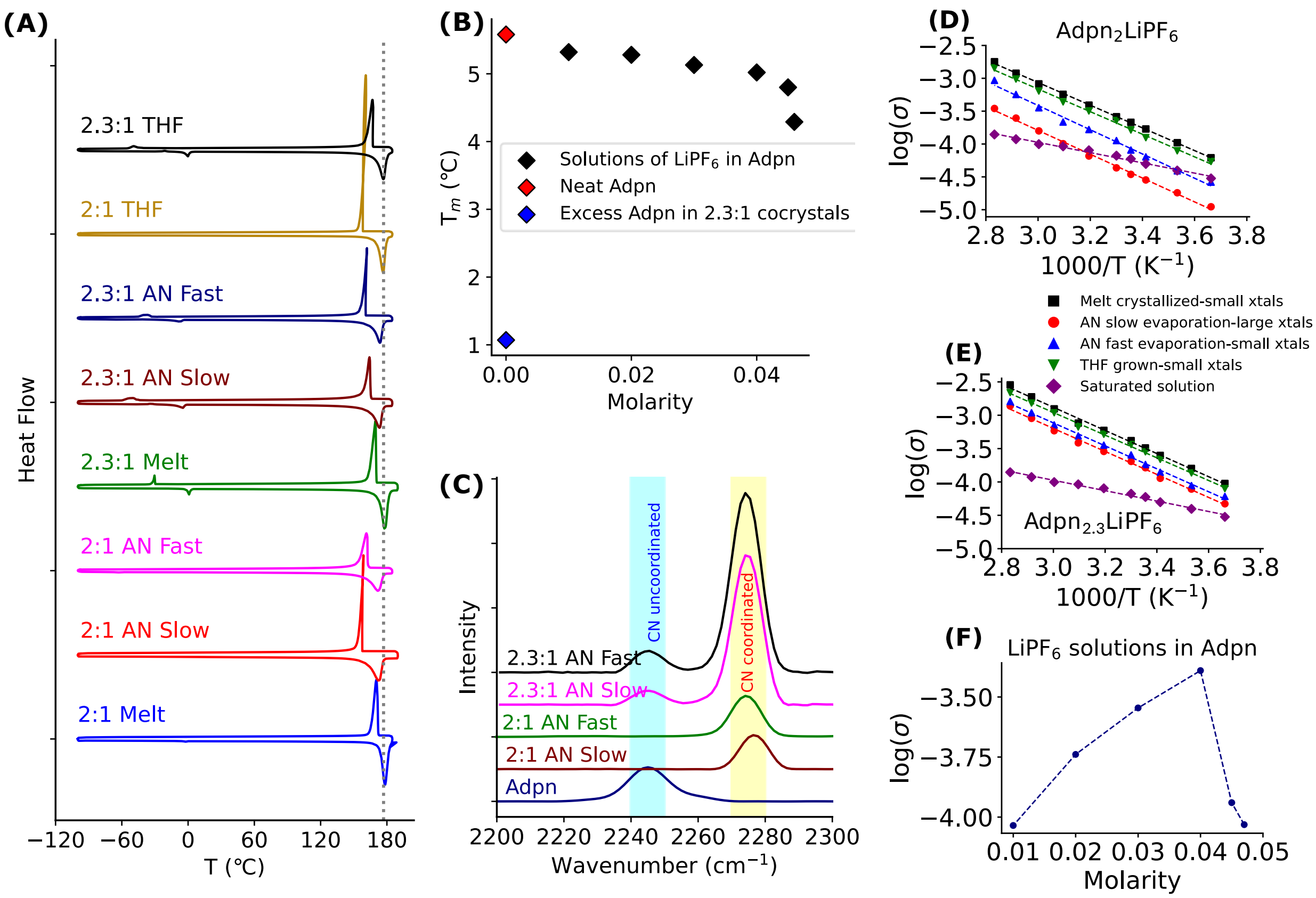

**Figure 2.** (**A**) DSC data (first cooling/second heating cycles) for $(Adpn)_2LiPF_6$ and $(Adpn)_{2.3}LiPF_6$ crystallized from tetrahydrofuran (THF), fast and slow crystallized from acetonitrile (AN), and melt crystallized. No transitions between -100 $^0$C and +100 $^0$C for $(Adpn)_2LiPF_6$ and small crystallization peaks for $(Adpn)_{2.3}LiPF_6$ around 0 $^0$C (**Table 1**). In all cases, corresponding enthalpies for melting ($\Delta H_{m2}$) and crystallization ($\Delta H_{c2}$) are same for each sample. (**B**) $T_m$ of neat Adpn (red), Adpn/$LiPF_6$ solutions up to the saturation limit at 25 $^0$C (black), and Adpn in the non-stoichiometric $(Adpn)_{2.3}LiPF_6$ (blue); (**C**) Raman spectra of Adpn, $(Adpn)_2LiPF_6$ and $(Adpn)_{2.3}LiPF_6$ stoichiometries of Adpn/$LiPF_6$, crystallized from AN by fast and slow evaporation. All $(Adpn)_2LiPF_6$ have a single peak at 2275 cm$^{-1}$; all $(Adpn)_{2.3}LiPF_6$ have peaks associated with neat and coordinated Adpn. The $\%\, free\ CN = \frac{I_{freeCN}}{I_{freeCN}+I_{bound}} = 13.0\%$. Conductivity data from heating cycles for: (**D**) Melt and solution crystallized $(Adpn)_2LiPF_6$, and saturated solution of $LiPF_6$ in Adpn; (**E**) Melt and solution crystallized $(Adpn)_{2.3}LiPF_6$, and a saturated solution of $LiPF_6$ in Adpn; (**F**) Solutions of $LiPF_6$ in Adpn from 0.01 M to solubility limit at 25 $^0$C. Activation energy barriers derived from Arrhenius fit of the datasets in (D) and (B) are provided in **Table S3** of the SI.

**Table 1.** Crystallization data derived from DSC thermograms shown in **Figure 2A**.Temperature (T), enthalpy (ΔH), subscripts: m melting, c crystallization, 1 corresponding to Adpn, 2 corresponding to molecular crystals.

| **Sample, crystallization method** | **Crystal size (μm)** | **(°C)** | | | | | **(J/g)** | | | |
|---|---|---|---|---|---|---|---|---|---|---|
| | | $T_{m1}$ | $T_{m2}$ | $T_{c1}$ | $T_{c2}$ | $\Delta T_{m2-c2}$ | $\Delta H_{m1}$ | $\Delta H_{m2}$ | $\Delta H_{c1}$ | $\Delta H_{c2}$ |
| **Adpn** | - | 3 | - | -25.2 | - | - | - | - | - | - |
| **$Adpn_2LiPF_6$, melt** | 10-25 | - | 178.1 | - | 170.1 | 8 | - | 175.2 | - | 177.7 |
| **$Adpn_2LiPF_6$, THF** | 25 | - | 177.0 | - | 162.0 | 15 | - | - | - | - |
| **$Adpn_2LiPF_6$, AN fast** | 25 | - | 173.0 | - | 159.3 | 14 | - | 172.6 | - | 171.3 |
| **$Adpn_2LiPF_6$, AN slow** | 100-200 | - | 173.1 | - | 158.8 | 14 | - | 172.7 | - | 171.8 |
| **$Adpn_{2.3}LiPF_6$, melt** | 25 | -1 | 178.0 | -30.4 | 169.0 | 9.2 | 15.7 | 174.2 | 14.5 | 176.3 |
| **$Adpn_{2.3}LiPF_6$, THF** | < 25 | - | 177.0 | - | - | - | - | - | - | - |
| **$Adpn_{2.3}LiPF_6$, AN fast** | 25 | -4.3 | 173.2 | -46.8 | 157.3 | 16 | 15.9 | 171.0 | 16.3 | 171.9 |
| **$Adpn_{2.3}LiPF_6$, AN slow** | 100-200 | -4.0 | 173.2 | -43.8 | 155.8 | 18 | 15.9 | 170.7 | 17.0 | 170.3 |

Excess Adpn in different samples, quantified from Raman spectra of "free" (2245 $cm^{-1}$) cyano (C≡N) groups of Adpn and those coordinated with $Li^+$ ions ( 2275 $cm^{-1}$ )[38] confirms that all the $(Adpn)_2LiPF_6$ with stoichiometric amounts of Adpn have a single peak at 2275 $cm^{-1}$ and no Adpn peaks, while all the crystals with excess Adpn, $(Adpn)_{2.3}LiPF_6$, have peaks associated with both the neat and the coordinated Adpn **(Figure 2C)**. Assuming, as was the case for acetonitrile[42], that the Raman scattering coefficients for the C≡N stretch is the same for the bound and free cyano groups, the % of free C≡N groups in the $(Adpn)_{2.3}LiPF_6$ samples is 13%, irrespective of their size.

Conductivity data for stoichiometric $(Adpn)_2LiPF_6$ (heating cycle **Figure 2D,** cooling cycle **Figure S6A**) is in the order: melt crystallized (10-25 μm) ≥ THF (25 μm) > AN fast crystallized (25 μm) > AN slow crystallized (200-300 μm). Differences between the $(Adpn)_2LiPF_6$ involve both the grains and the GBs. The conductivity trends support these effects: (**i**) Smaller crystals have higher conductivity than larger crystals since their higher surface areas have higher GB contributions; (**ii**) Small melt crystallized $(Adpn)_2LiPF_6$ has higher conductivity than the small solution crystallized $(Adpn)_2LiPF_6$, since GBs fuse during the melt crystallization process; (**iii**) Defects in the grains (not GBs) decrease conductivity: σ (small melt and THF solution crystallized $(Adpn)_2LiPF_6$ without defects) > σ (small AN solution crystallized $(Adpn)_2LiPF_6$); immobile and low mobility point defect obstructions were previously shown to decrease diffusion constants of $Li^+$ ions.[43] Although AN could also reside in the GBs, this would result in higher conductivity than for crystals made using THF or melt crystallization, since the conductivity of $LiPF_6$ in AN ($\sigma \sim 5 \times 10^{-2}$ S/cm, 1M $LiPF_6$ in AN at 25 $^0$C)[44] would, but does not, enhance the conductivity of the AN solution crystallized samples.

For the non-stoichiometric $(Adpn)_{2.3}LiPF_6$, the same conductivity trends (**Figure 2E**, **Figure S6B**) are observed. Further, excess Adpn increases the conductivity of all non-stoichiometrically prepared $(Adpn)_{2.3}LiPF_6$ crystals compared with the corresponding stoichiometrically prepared $(Adpn)_2LiPF_6$, suggesting that the Adpn resides between the grains and increases $Li^+$ ion diffusion. To understand these results, the conductivity of dilute solutions of $LiPF_6$ in Adpn was measured (**Figure 4C**) and peaked at 0.046 M, ≈ two orders of magnitude lower in concentration than the ~ 1M typically observed for Li salts in aprotic solvent, due to the low solubility of $LiPF_6$ in Adpn. A saturated solution of $LiPF_6$ in Adpn has a conductivity lower than all the non-stoichiometric $(Adpn)_{2.3}LiPF_6$ and stoichiometric $(Adpn)_2LiPF_6$ cocrystals (except for the AN slow crystallized sample at T < 50 $^0$C). This suggests that the enhanced conductivity for the non-stoichiometric $(Adpn)_{2.3}LiPF_6$ cocrystals is not due to "channels" of dilute $LiPF_6$/Adpn solutions between the crystal grains, which would lead to decreased ionic conductivity.

Further, the activation energies ($E_a$) from temperature dependent ionic conductivities (**Table S3**) are almost identical for the stoichiometric (also has GBs) and the non-stoichiometric crystal that contains excess solvent in the GBs. $E_a$ for both stoichiometric and nonstoichiometric molecular crystals ($\Delta E_a \approx 33.5$ kJ/mol) is twice ($\Delta E_a \approx 14.9$ kJ/mol) that of the 0.04M $LiPF_6$ in Adpn, a $\Delta E_a$ typical for LiX in carbonates/ethers ($\Delta E_a \approx 8$–$15$ kJ/mol). These results, along with the melting point data (**Figure 2B**), strongly suggest that GBs are complex interfacial regions where ions move in a nanoconfined space between the grains and can be swollen with excess Adpn. The GB region is more conductive than would be the case for a "channel" of low conductivity 0.04 M $LiPF_6$ in Adpn liquid, but less conductive than for a channel of 1M Adpn if it were soluble at this molarity.

The presence of excess Adpn in the GB region minimizes the difference between the AN solution crystallized small and large crystallized samples) $(Adpn)_{2.3}LiPF_6$ since it serves to better wet and connect the crystal grains, minimizing differences arising from differences in surface area. Other nitriles- succinonitrile (SN) and 1,3,6-hexanetricarbonitrile (HTCN), which are excluded from the crystal lattice to the GBs, also increase the ionic conductivity (**Figure S7A, S7B**), but can decrease conductivity when viscosity increases.[45] The fluid GB regions make it possible to swell these regions with polymer or polymer gels, decreasing/eliminating the interfacial resistance in polymer/polymer gel and molecular crystal composites.[46]

The effect of grain size on the concentration and dynamics of charge carriers in nano-confined environments can be modeled using molecular dynamics (MD) simulations. The contribution of grain-boundaries (GB) and their thickness to ionic conduction have been shown earlier in a similar class of electrolytes, using MD simulations, where the $Li^+$ ion motion in the grains was characterized by a sub-diffusive hopping mechanism 100 times slower than in the GBs. [38, 47] In the current simulations, an atomistic classical force-field[38 47] that was modified to predict diffusion coefficients and transference numbers of $Li^+$ ions in pristine $(Adpn)_2LiPF_6$ and molecular crystals with an excess solvent environment was used. Previous models were extended to increase the grain size (**Figure S8a, b**) and probe the diffusion of ions from different 010 and 001 crystal facets (**Figure S8c, d**). The initial excess volume of Adpn (a minimum of 1nm on each side) was less than would be needed for the 2.4:1 stoichiometry, which is inaccessible at this scale due to size limits. **Figure 3A-C** shows the trajectory of $Li^+$ ions in these three models during the timescale of the simulation. **Figure 3A** shows solvation of $Li^+$ ions at the edges and vertices more than for the 010 or 001 surface, suggesting these are important contact points for multiple grains to form viable GB regions. This is consistent with the rounded edges observed in the SEM images of the crystals (**Figure 1**). **Figure 3B** shows $Li^+$ trajectories for a supercell exposed in the *b*-crystallographic direction (010 surface). It is important to note that $Li^+$ ions in this channel were previously shown to have the lowest energy barrier for migration.[38] This is due to a 6.2 Å successive distance between $Li^+$ ions in this, '*b*', crystallographic direction, no CIP formation during the transition state (observed from DFT),[38] and the ability to enter/exit the GBs at both ends. **Figure 3C** shows that in the *c*-crystallographic (001) direction, $Li^+$ ions are not solvated (left-inset) unless the initial coordination layer with Adpn molecules is half (right-inset). The initial excess solvent layer (2.5 nm) does not populate with charge carriers as well as in the case of 010 surface. The results for the *c*-crystallographic direction can also be applied to the "*a*" crystallographic direction (100 surface), suggesting that only the *b*-crystallographic direction contributes effectively to the grain boundary conduction.

The concentrations of charge carriers in the GB regions are computed using a number density distribution for $Li^+$ ions in each cross-sectional surface slab **(Figure 3D,E)**. It is useful to consider the GBs as consisting of regions directly adjacent to the grains (interfacial region) and the region further from this interface, referred to as GB-solvated regions. The number density distribution shows that for the 001-surface model **(Figure 3D)**, the number of charge carriers remain roughly the same near the start (0 – 5 ns) of the simulation and near the end (45 – 50 ns) of the simulation. For the 010 surface, where most of the $Li^+$ ions enter the GB region, a significant continuous growth in the $Li^+$ number density was observed over the simulation time **(Figure 3E).** The equilibrium $Li^+$ ion concentration (calculated from 45 – 50 ns window) in the GB regions of 010 model was significantly (~ 25 times) larger (~25 times, ~ 0.5 $nm^{-3}$, ~ 1 M) than in the supersaturated solution of $LiPF_6$ in Adpn (0.04 M). This predicted concentration (1 M) of $Li^+$ ions in the nanoscale GB regions is a similar order of magnitude to its value predicted from extrapolation of the melting point suppression plot (**Figure 2B,** 0.4 M).

To assess the ionic mobility in the aforementioned model structures, mean-squared displacements (MSDs) of the $Li^+$ and $PF_6^-$ ions were calculated for each individual entity (**Figure S9**). The MSD *vs.* time plots show that there are more linearly diffusive ("well-diffusive") charge-carriers in the 1g model than in the 010 and 001 surface models. The distinction further confirms the role of vertices and edges forming fluid grain-boundary regions in the 1g model. The well-diffusive charge carriers are further filtered (where the standard deviation is < 1 % in their individual slopes for multiple time origins of the MSD *vs.* time plots) to obtain distribution histograms of diffusion coefficients for each model (**Figure 4A-F**). Roughly 10 % (out of 2048 total $Li^+$ and $PF_6^-$ each) of these charge carriers contribute to ion-conduction as diffusive charge carriers in the 1g and 010 grain models. This number drops significantly to only ~4 % for the 001 grain surface model case. In comparison with the earlier investigated smaller sized (higher surface/volume ratio) 'two-grain' model[38] (where roughly 40 % out of 1000 $Li^+$ and $PF_6^-$ ions each are located in the grain boundaries), the well-diffusive charge carriers are less for the larger sized (smaller surface/volume ratio) 1g model (only 10 %). This agrees with the experimentally observed effect of grain-size, where crystals with smaller grains have higher ionic conductivity than those with bigger grains **(Figure 2E)** due to the increase in charge carriers in the GB for the smaller crystals. The ratio of ionic conductivities for small:big grains is ~ 8:1 from experimental data **(Figure 2D, Figure S6A)**, compared with 8:5 for the 2g (2 small grains):1g (1 big grain) models from simulation data. We can certainly assume that if the concentration of nano-confined charge carriers changed from 10 % to 40 % for an 8:5 size ratio in simulations, in experimental crystals an 8:1 difference can potentially boost the ionic conductivity by two orders of magnitude **(Figure 2E).**

The average diffusion coefficients ($D_{Li^+}$ and $D_{PF_6^-}$), calculated for all ions and then separately for interfacial and GB-solvated regions **(Table 2)**, show that the GB solvated ions are ~ one order more mobile compared to interfacial ions, in all three models. The average diffusion coefficients for the 010 ions are largest, reaffirming it to be the most potent GB forming facet of the cocrystals. Further, only slightly higher coefficients for the 1g model compared to 001 model are due to the presence of four (two 100 and two 001) low-conductive surfaces out of six, and not enough thickness for the excess solvent layers.

We further characterized the nature of $Li^+$ (and $PF_6^-$) ions in the GB regions by calculating interatomic interactions of $Li^+$ ions as radial distribution functions (RDFs) with F($PF_6^-$) and N(Adpn) for both the grain and GB regions (**Figure 4G)**. In the grains at short distances (~ 0.2-0.3 nm), there are negligible $Li^+\cdots F(PF_6)$ interactions, consistent with the crystallographic data. The calculated coordination numbers at first minima for the grains show negligible contact ion pair (CIP) formation. In the GB region, there are more CIPs observed. The calculated coordination numbers at first minima show ~ 0.2 ion-pairs forming per $Li^+$ ion in the GB regions. To further identify if these ion-pairs are long-lived or short-lived, the F atom neighbors around $Li^+$ ions **(Figure 4H)** are calculated where ~ 210 single Li···F bonded short-lived ion-pairs are seen with a cut-off of 3.0 Å. Stronger, long-lived ion-pairs (with two Li···F bonds) are observed in only a very small number (~ 2 – 3). Further for GB $Li^+$ ions, a very similar (to grain) Li…N RDF and coordination number suggests that the $Li^+$ ions in the GB region are predominantly solvated by Adpn. Both these results indicate that most of the charge-carriers in the GB regions are not in CIPs and thus contribute to ionic conductivity without any cross-correlations.

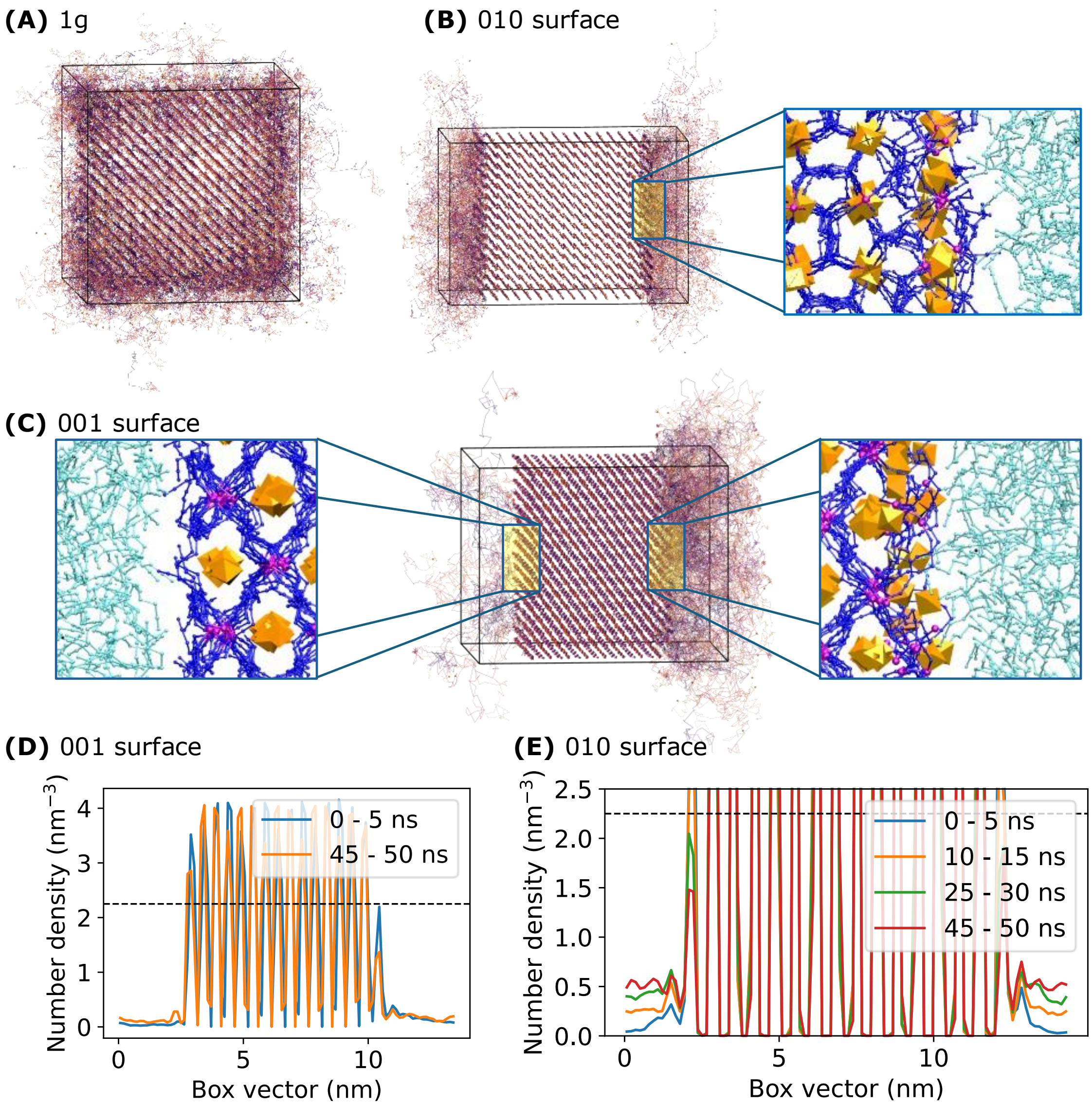

**Figure 3.** Trajectory of Li+ ions in: **(A)** a single large grain (1g), which shows solvation and then exchange of Li+ and PF6- ions from all sides of the crystal, while edges and vertices solvate more justifying a rounded (soft-edges) appearance of crystals in SEM images; **(B)** the 010 surface and **(C)** the 001 models of the grain-boundary. The trajectory lines show movement of Li+ and PF66- ions from t = 0 (dark) to t = 50 ns (light) lines sampled at every 0.5 ns. For the 010 and 001 surfaces, a zoomed-in view of interface is shown (inset), providing a comparative view of surface termination with inherent Adpn (blue balls and sticks, part of crystal stoichiometry) vs. excess Adpn (cyan balls and sticks, external solvent molecules). The 010 surface shows consistent solvation and exchange of charge carriers in the GB region. For the 001 surface, only the ions on the right side are solvated and exchanged due to incomplete initial coordination of Li+ ions, the left side remains intact indicating that completely solvated Li+ ions in the c-crystallographic direction do not contribute significantly in generating long-term charge carriers. Number density of $Li^+$ ions across the box in the direction perpendicular to the exposed surface for the **(D)** 001 surface, and **(E)** 010 surface.

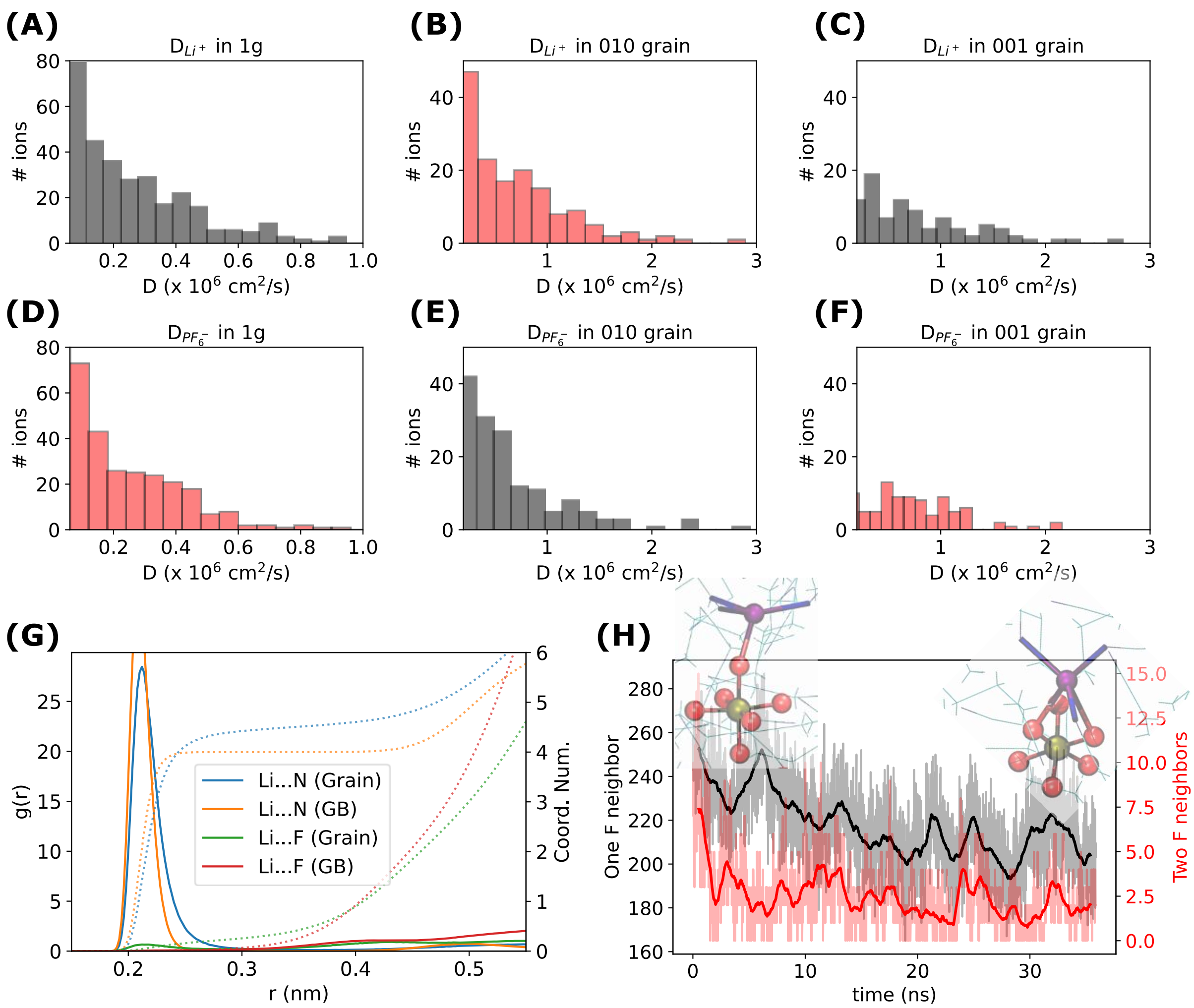

**Figure 4.** Distribution of self-diffusion coefficients for $Li^+$ ions in **(A)** 1g, **(B)** 010 grain, **(C)** 001 grain, and for $PF_6^-$ ions in **(D)** 1g, **(E)** 010 grain, **(F)** 001 grain, only for the charge-carriers which are "well-diffusive", defined using slopes MSD *vs.* time plots for each individual charge carrier in Figure S9; **(G)** Li…N(Adpn) and Li…F($PF_6^-$) radial distribution functions for ions quantified as "in grains" and "in GB", with coordination numbers for $Li^+$ ions with N/F atoms in GB and in grains, calculated from RDFs, particularly showing negligible cation-anion interactions in GB. **(H)** Number of $Li^+$ ions with one (plateaued to ~ 180) or two (plateaued to ~5) F ($PF_6^-$) neighbors in the GB regions. Left inset image shows a sample geometry for short-lived Li…$PF_6$ ion-pair in GB regions, where a $Li^+$ ion (purple) is seen coordinating with one F atom of $PF_6^-$ (red) and three N atoms (blue) from three Adpn molecules. Right inset shows a sample geometry for a longer-lived Li…$PF_6$ ion-pair in GB regions, where a $Li^+$ ion is seen coordinating with two F atoms of a single $PF_6^-$ and three N atoms from three Adpn molecules.

**Table 2.** Average diffusion coefficients for $Li^+$ and $PF_6^-$ ions, $D_{Li}^+$ and $D_{PF6}^-$ for all well-diffusive ions, and their averages in interfacial and GB solvated regions. For all the well-diffusive ions, the average diffusion coefficients are divided in two boundary limits of $D_{interfacial} < 0.8 \times 10^{-6}$ cm²/s < $D_{GB\text{-}solvated}$. The interfacial ions, between GB and grains, do not travel far enough in the GB to be a part of nanoconfined region, and are also in rapid exchange with the crystal interior bound Adpn molecules. Ions with D > 0.8 $\times 10^{-6}$ cm²/s are the GB solvated ions, which are part of the nanoconfined solvated region.

| | | **D (x 10⁶ in the units of cm²/s)** | | | | | |
|---|---|---|---|---|---|---|---|
| **Model** | | **All Ions** | | **Interfacial ions** | | **GB solvated ions** | |
| | | Average | Std.dev | Average | Std.dev | Average | Std.dev |
| **1g** | $Li^+$ | 0.06 | 0.14 | 0.06 | 0.13 | 0.92 | 0.09 |
| | $PF_6^-$ | 0.05 | 0.12 | 0.05 | 0.13 | 0.92 | 0.1 |
| **010** | $Li^+$ | 0.36 | 0.51 | 0.17 | 0.23 | 1.29 | 0.51 |
| | $PF_6^-$ | 0.23 | 0.43 | 0.12 | 0.2 | 1.36 | 0.56 |
| **001** | $Li^+$ | 0.05 | 0.23 | 0.02 | 0.1 | 1.36 | 0.42 |
| | $PF_6^-$ | 0.04 | 0.19 | 0.01 | 0.09 | 1.19 | 0.33 |

Soft-Solid molecular crystals consist of grains and fluid grain boundaries that enhance transport of $Li^+$ ions between the grains. Unlike LICC where (due to resistive grain boundaries) pressure or sintering is required to improve contact between the grains, the soft grain boundaries in molecular crystals can be connected by pressure, melting, or addition of a slight excess of the organic component between the grains. Both experimental ionic conductivities for different sized and solvated samples and MD simulations show that the total conductivity consists of $Li^+$ ion migration in the grains as well as in the grain boundaries. Since the grain boundaries can be swollen by other low molar mass compounds (e.g. other dinitriles) and/or the fluid GBs can swell a polymer, there is no interfacial resistance between the two components (molecular crystals and polymer or polymer gel). More factors like solvent polarizability, solvent dielectric constant, and ion-pair stabilization energies are to be explored to tailor the GB regions for optimal stability and conductivity.

## Experimental

### Materials

The adiponitrile (Adpn), acetonitrile (AN), tetrahydrofuran (THF) and $LiPF_6$ salt were purchased from Sigma-Aldrich. The Adpn was degassed using a Schlenk line and stored in a vial with molecular sieves until used in an Ar purged glove box. The AN and THF were also distilled and degassed before bringing them into the glovebox. Dimethyl sulfoxide-$D_6$ (D, 99.9%) was purchased from Cambridge Isotope Laboratories. All syntheses were performed in an Ar purged glove box. Samples for DSC and electrochemical analysis were prepared in hermetic pans or coin cells, respectively, in an Ar purged glove box.

**Cocrystal and solution preparation**

Crystalline materials of adiponitrile (Adpn) and $LiPF_6$ were made with stoichiometric ratios (Adpn)/$LiPF_6$ of 2/1, $(Adpn)_2/LiPF_6$, and 2.3/1, $(Adpn)_{2.3}/LiPF_6$. The crystals were prepared either by melt or solution crystallization. The melt crystallized samples were prepared by heating the two components until the $LiPF_6$ dissolved in the Adpn (~ 180 °C) and then letting the solution cool to room temperature. Crystallization occurred almost immediately, i.e. at high temperatures (close to $T_m$), as is evident from the DSC data below. Due to the small difference between $T_m$ and $T_c$, it was not possible to grow larger crystals by holding the sample between these two temperatures for long times. Solution crystallized samples were prepared by dissolving Adpn and $LiPF_6$ at the appropriate ratios in the cosolvents, acetonitrile (AN) or tetrahydrofuran (THF), and evaporating the AN cosolvents. By changing the rate of evaporation, it was possible to grow both large and small crystals. The solution crystallized samples were "fluffier" in appearance than those prepared from the melt. To investigate the conductivity of solutions of $LiPF_6$ in Adpn, solutions of $LiPF_6$ in Adpn were prepared by dissolution of the salt in the solvent until the solutions were saturated and precipitation was observed.

**Characterization**

**Thermogravimetric Analysis (TGA)** data was acquired on a TA Instruments Hi-Res TGA 550 at a ramp rate of 10 °C $min^{-1}$ with a flow of ultra-pure N2 gas and was used to determine weight percent loss of Adpn. Differential scanning calorimetry (DSC) on a TA Instruments Model 250 DSC was used to determine melt temperatures ($T_m$) and enthalpies ($\Delta H_m$) during the second heating cycle and crystallization temperatures ($T_c$) on the first cooling cycle. Samples in hermetically sealed Tzero aluminum pans were scanned from -100 °C to 265 °C (or until just above $T_m$ of the crystals) at a scan rate of 10 °C $min^{1}$, under ultra-pure N2 purge.

**X-ray diffraction (XRD)** data was obtained on a Bruker Kappa APEX II DUO diffractometer. Single-Crystal X-ray Diffraction (SCXRD) was obtained using Mo-Kα radiation. Data was collected at 100K. Data were processed using the Bruker Suite. The structure was solved and refined using the SHELXTL package[48] with Olex2[49] as a GUI. Powder X-ray diffraction (PXRD) data was obtained using Cu-Kα radiation. Data were collected at 100 K. The calculated powder pattern was generated from the single crystal $(Adpn)_2LiPF_6$) data using MercuryCSD.[50] The simulated data conformed with the experimental powder data.

**SEM data** was acquired on an FEI Quanta 450FEG (field emission gun) SEM instrument.

**NMR data** was collected on Bruker AVIIIHD 500 by dissolving the cocrystals in deuterated dimethyl sulfoxide ($d_6$-DMSO).

**Raman spectra** were recorded in the 100-3000 $cm^{-1}$ region at room temperature using a Horiba LabRAM HR Evolution Raman spectrometer, with a resolution of 1.8 $cm^{-1}$, an excitation wavelength of 532 nm, 60 mW power, and a grating groove density of 600 gr/mm. Samples were measured with 8 acquisitions, and 2 to 8 seconds each, depending on peak intensity.

**Conductivity** measurements were obtained on press pellets of $(Adpn)_2LiPF_6$ or $(Adpn)_{2.3}LiPF_6$ in 2032-type coin cells by AC electrochemical impedance spectroscopy (EIS). The coin cells were made by compressing polycrystalline $(Adpn)_2LiPF_6$ or $(Adpn)_{2.3}LiPF_6$ powder in an O ring shaped Teflon separator between two stainless steel spacers, thus maintaining a consistent thickness from sample to sample. A glass fiber separator (0.20-0.25 mm) was used for measuring the conductivity of a saturated solution of $LiPF_6$ in Adpn. Temperature-dependent bulk impedance data were acquired using a.c. electrochemical impedance spectroscopy (EIS) by synchronizing a Gamry Interface 1000 potentiostat/galvanostat/zero-resistance-ammeter (ZRA) in the frequency range 0.1−1 MHz) with an ESPEC benchtop chamber (Model-BTU-133) in a temperature range between $0^0$ and 80 °C with increments of 10 °C. The cell was thermally equilibrated for 30 minutes at each temperature before the impedance was measured during both the cooling and heating cycles. Control of the equipment was through Gamry framework software and data was analyzed with Gamry Echem analysis software. The ionic conductivity (σ) was calculated using equation (1):

$$\sigma = \frac{t}{RA} \qquad (1)$$

where t is the thickness (cm), R is the volume resistance (Ω), and A is the area of the electrolyte in contact with the electrode ($cm^2$). Activation energies ($E_a$) were calculated using the Arrhenius equation (2):

$$\sigma = \frac{A}{T} exp\left(\frac{-E_a}{k_b T}\right) \qquad (2)$$

where σ is the ionic conductivity, A is the prefactor, related to the number of charge carriers, $k_b$ is the Boltzmann constant, T is the absolute temperature.

**MD Simulations**

The effect of grain size on the concentration and dynamics of charge carriers in nano-confined environments can be modeled using molecular dynamics (MD) simulations. The presence of grain boundaries in soft-solid crystals has been observed to contribute prominently to the net ion-conduction.[38,47] Molecular dynamics was used to investigate and better understand the concentration and mobility of $Li^+$ ions in polycrystalline Adpn:$LiPF_6$ complexes. The application of MD simulations to estimate the equilibrium dimensions of the grain boundaries in Adpn:$LiPF_6$ complexes would not be practical, as these regions exist at µm sizescales and would require µs to ms long timescales. In the current simulation an atomistic classical force-field[38 47] that was modified to predict diffusion coefficients and transference numbers of $Li^+$ ions in pristine $(Adpn)_2LiPF_6$ and molecular crystals with an excess solvent environment was used. Four atomistic models are used to compare the effect of grain size on the nature of grain boundaries and ion conduction. The grain-boundary effects were previously modeled using two grains (2g) of 5x5x5 unit cells that were solvated in an excess of 6286 Adpn molecules **(Figure S1a)**. In the current investigation we implemented important improvements to this model: a larger single grain (1g) constructed of 8x8x8 unit cells of $(Adpn)_2LiPF_6$ (2048 $Li^+PF_6^-$, 4096 Adpn – a total of 8192 atoms) with an initial excess volume of a minimum of 1nm on each side **(Figure S1b)**. This supercell was then solvated at the grain boundaries in a cubic periodic box consisting of 3080 excess Apdn molecules, producing an effective ratio of 5:1 for Adpn: $LiPF_6$. It is important to note here that experimental stoichiometries with excess solvent, 2.4:1 used here, are inaccessible at this scale due to size limits. The third and fourth models are semi-isotropically solvated with excess Adpn molecules from the *b* (010) and *c* (001) crystallographic directions (**Figure S1c, d**). These models **(Figure 1b, c, d)** were first equilibrated for 20 ns under *NpT* ensemble conditions at 300 K and 1 bar pressure using a velocity rescale thermostat[51] and Berendsen barostat.[52] A minimum 50 ns long simulation trajectory was then acquired for each of these models to compute mean-squared displacements, concentration of charge carriers, etc. All the simulations were performed with Gromacs 5.0.7 software package[53] with a uniform timescale of 1.0 fs and a cut-off of 14 Å for the computation of Coulombic and vdW forces.

**PXRD and SC-XRD**

PXRD data (**Figure S1**) show excellent agreement between the single crystal XRD of $(Adpn)_2LiPF_6$[38] and all the samples: stoichiometric $(Adpn)_2LiPF_6$ compositions or compositions with excess Adpn, $(Adpn)_{2.3}LiPF_6$, either prepared by melt (reported earlier[38]) or solution crystallization. No evidence of Adpn peaks occur in the PXRD data of $(Adpn)_{2.3}LiPF_6$. The PXRD samples were rapidly quenched to 100K, where the PXRD data was acquired. In this case, the liquid layer of $LiPF_6$ in Adpn formed a glass with no powder pattern. When slowly cooled (for DSC runs, see below) the liquid layer of $LiPF_6$ in Adpn crystallized near 0 $^0$C, and so was a liquid for the conductivity experiments.

## AUTHOR INFORMATION

### Corresponding Authors

* Michael J. Zdilla, Email: michael.zdilla@temple.edu

* Prabhat Prakash, Email: pprakash@caltech.edu

* Stephanie L. Wunder, Email: slwunder@temple.edu

### Author Contributions

The manuscript was written through contributions of all authors. All authors have given approval to the final version of the manuscript.

## ACKNOWLEDGMENT

This work is supported by National Science Foundation Grant (Award No. DMR 2138432), NSF MRI grant (CHE-2215854) and a Department of Energy Grant (Award No. DE-SC0023356). W.A.G. acknowledges support from the Hong Kong Quantum AI Lab, AIR@ InnoHK of the Hong

Kong Government. W.A.G. thanks the U.S. National Science Foundation (CBET- 2311117) for support.

**Declaration of Interests**

The authors have no competing interests to declare.

**Supplementary Material**

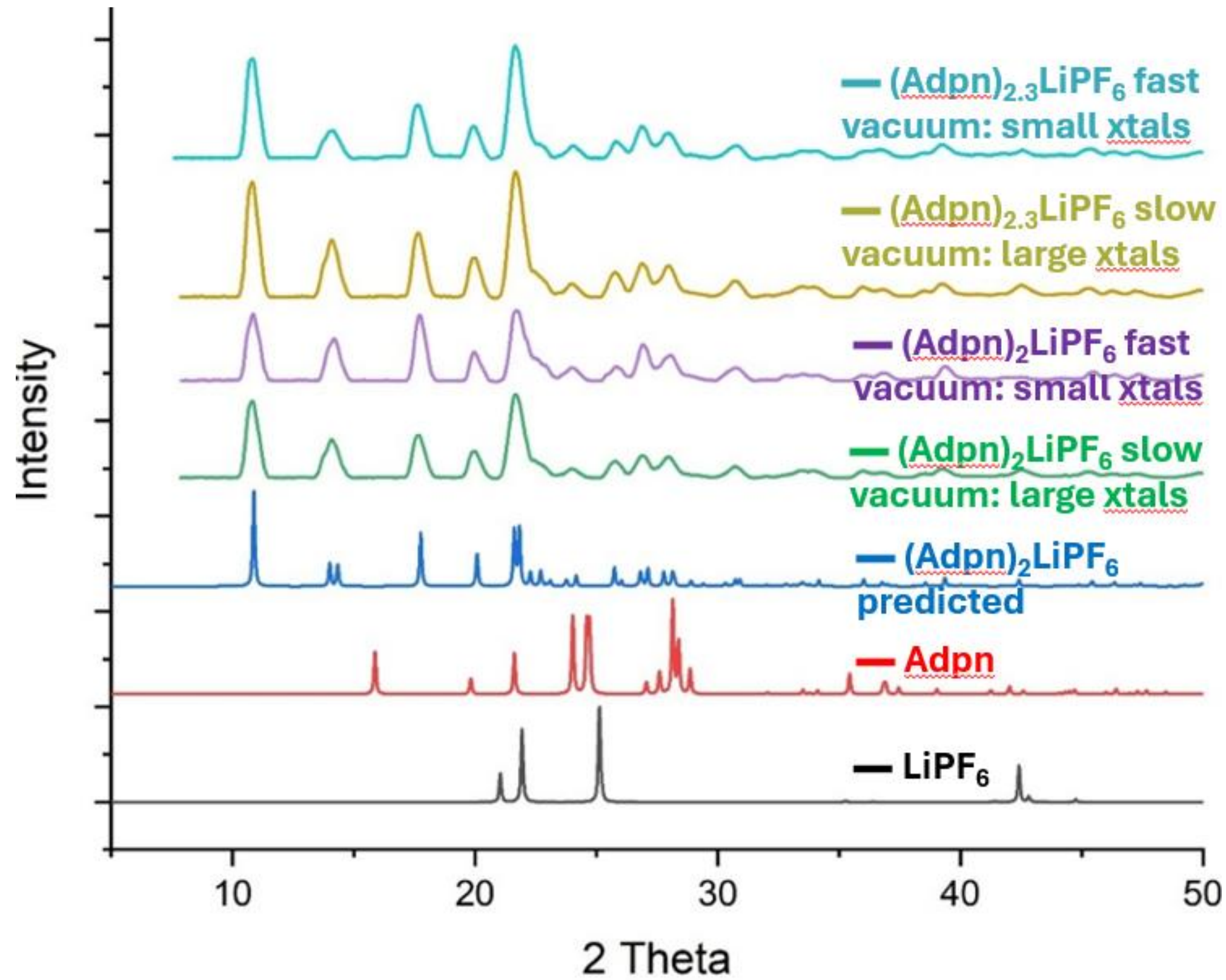

**Figure S1**. XRD data of starting materials adiponitrile (Adpn) and $LiPF_6$, predicted powder pattern of $(Adpn)_2LiPF_6$ from single crystal data, and PXRD of cocrystals $(Adpn)_{2.3}LiPF_6$ and $(Adpn)_{2.3}LiPF_6$ prepared by slow and fast evaporation from acetonitrile (AN).

**Crystal Sizes**

Since the purpose of the experiments was to obtain size variations in the crystals, many variables such as time to crystallization and solvent were explored. As expected, crystal size increased with increased time to crystallize. Slow evaporation (~6h) of acetonitrile (AN) yielded larger crystals than crystals formed by fast evaporation (~30 min), and melt crystallization was the fastest (< 5 minutes) resulting in small crystals. SEM images (**Figure 2**) of the crystals grown by these methods confirm that the melt and fast solution (AN) grown crystals are comparable and smaller in size than the slow (AN) solution grown crystals. Sizes for both $(Adpn)_2LiPF_6$ and $(Adpn)_{2.3}LiPF_6$ were in the order: melt crystallized (~ 10 µm) < fast crystallization from AN (~ 25 µm) < slow crystallization from AN (200 µm -300 µm length). The $(Adpn)_2LiPF_6$ crystallized from AN maintained its size after being melted in the DSC to just above its $T_m$ (**Figure S2**). $(Adpn)_2LiPF_6$ crystallized/evaporated from THF were also small ~ 25 µm (**Figure S3**)

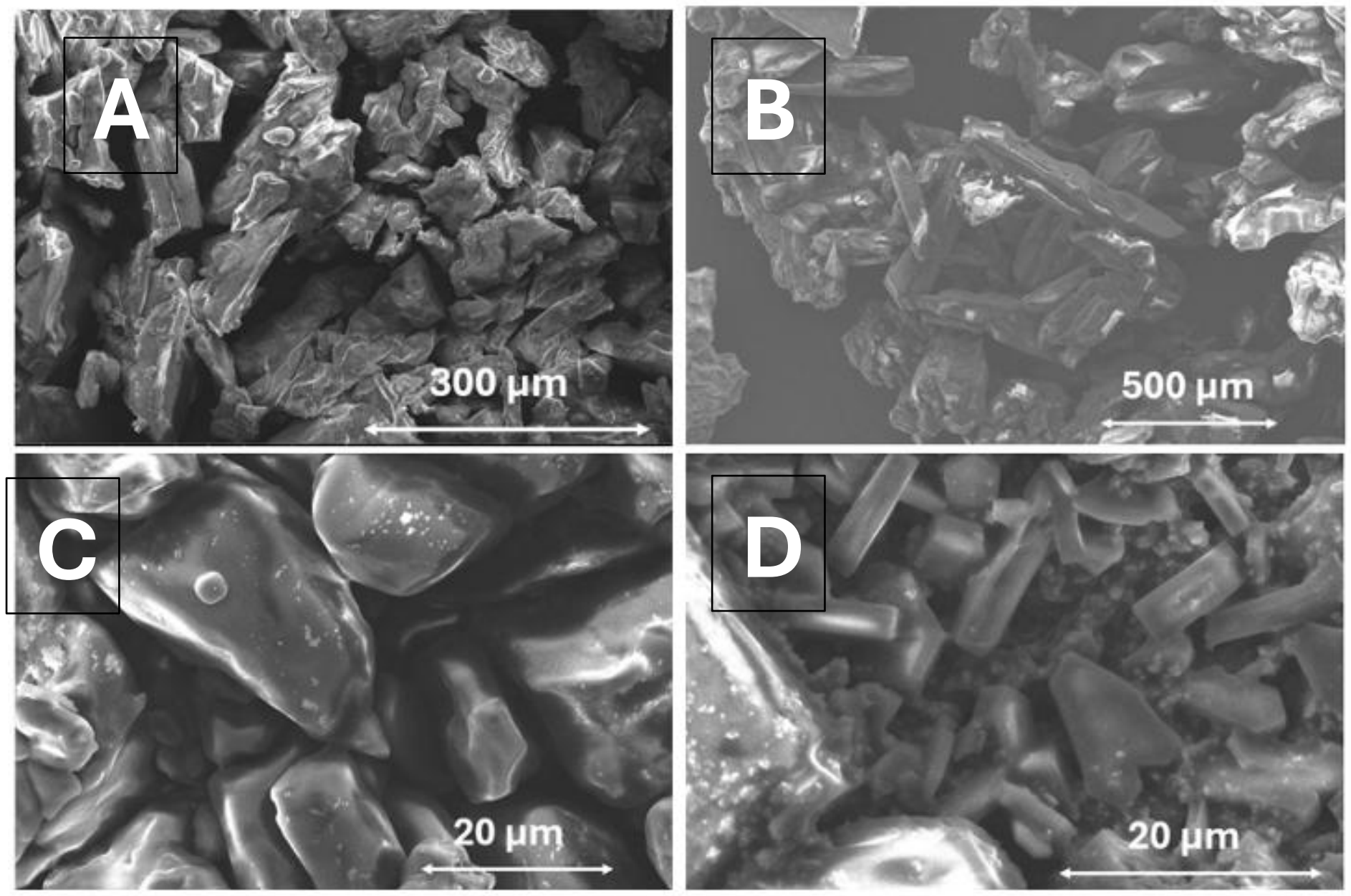

**Figure S2**. SEM images of $(Adpn)_2LiPF_6$ (**A**) crystallized slowly from acetonitrile (AN); (**B**) after heating sample "**A**" above $T_m$ in the DSC; (**C**) crystallized fast from acetonitrile; and (**D**) after heating sample "**C**" above $T_m$ in the DSC.

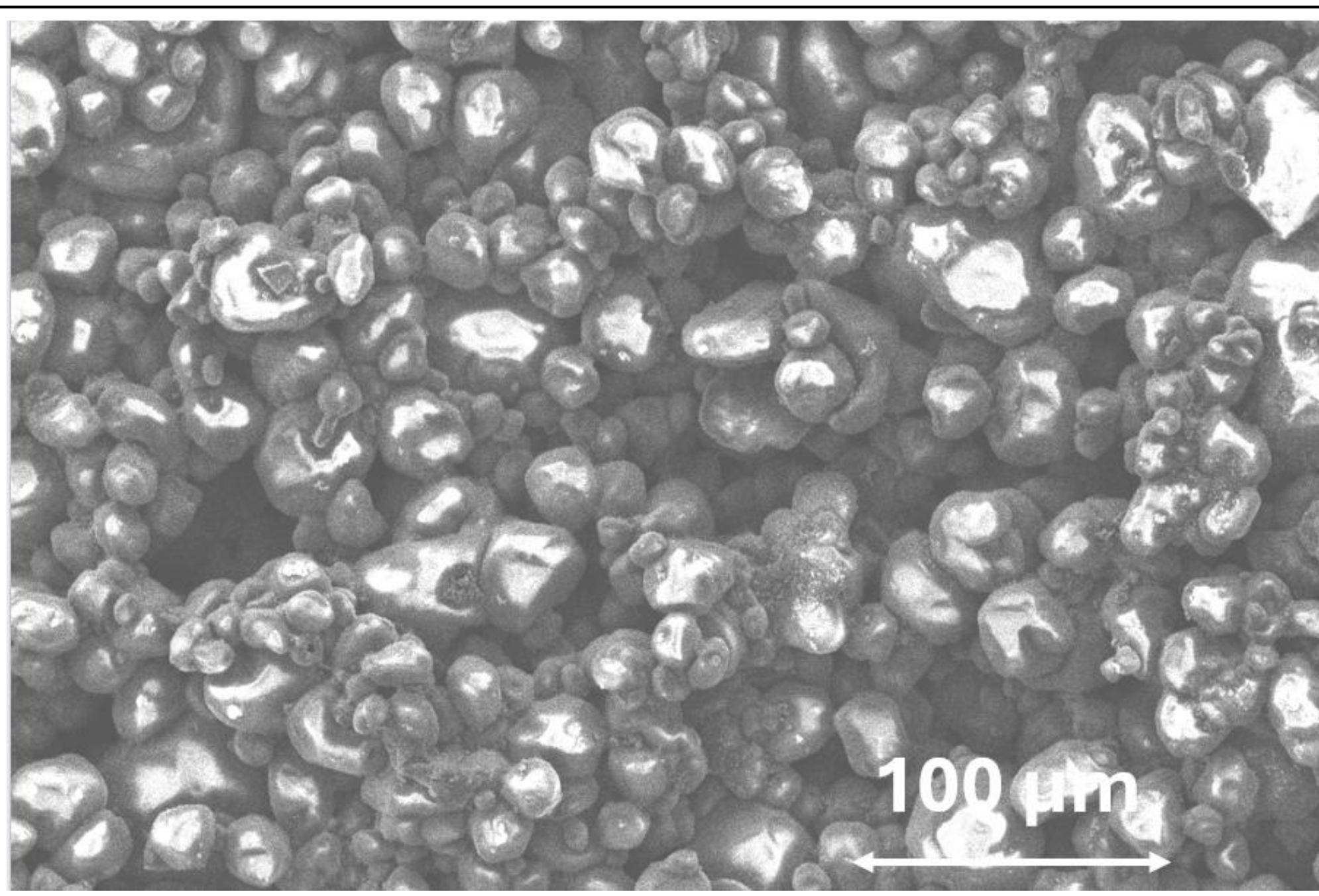

**Figure S3**. SEM image of $(Adpn)_2LiPF_6$ crystallized from tetrahydrofuran (THF).

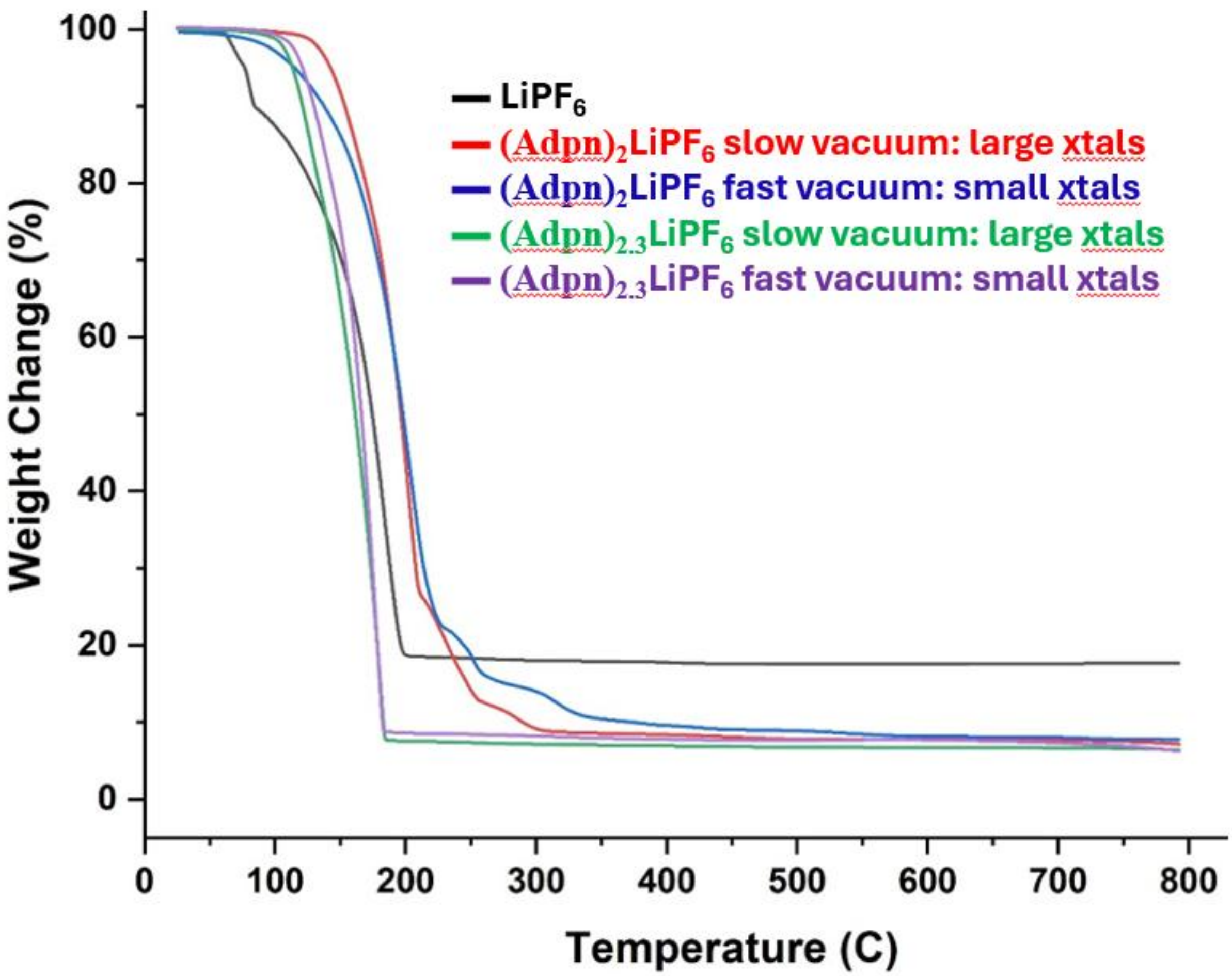

**Figure S4**. Thermogravimetric analysis (TGA) data for $LiPF_6$, $(Adpn)_2LiPF_6$ and $Adpn)_{2.3}LiPF_6$ molecular crystals prepared by evaporation from acetonitrile (fast and slow). Since $LiPF_6$ has 17.76% residual mass remaining at 800 ºC, the additional weight loss can be attributed to Adpn, which has no residual weight loss at 800 $^0$C (not shown).

**Table S1.** Remaining wt. % after removal of Adpn from calculated from TGA data and theoretical estimates.

| Sample | Xtal Size | Weight % Remaining |
|---|---|---|
| **Adpn** | | 17.70 |
| **$(Adpn)_2LiPF_6$ theoretical** | - | **7.30** |
| **$(Adpn)_2LiPF_6$ AN slow** | Large | 7.33 |
| **$(Adpn)_2LiPF_6$ AN slow** | Large | 7.01 |
| **$(Adpn)_2LiPF_6$ AN fast** | Small | 7.26 |
| **$(Adpn)_2LiPF_6$ Melt** | Small | 7.26 |
| **$(Adpn)_{2.3}LiPF_6$ theoretical** | - | **6.44** |
| **$(Adpn)_{2.3}LiPF_6$ AN slow** | Large | 6.41 |
| **$(Adpn)_{2.3}LiPF_6$ AN slow** | Large | 6.14 |
| **$(Adpn)_{2.3}LiPF_6$ AN fast** | Small | 6.39 |
| **$(Adpn)_{2.3}LiPF_6$ Melt** | Small | 6.24 |

**(A)**

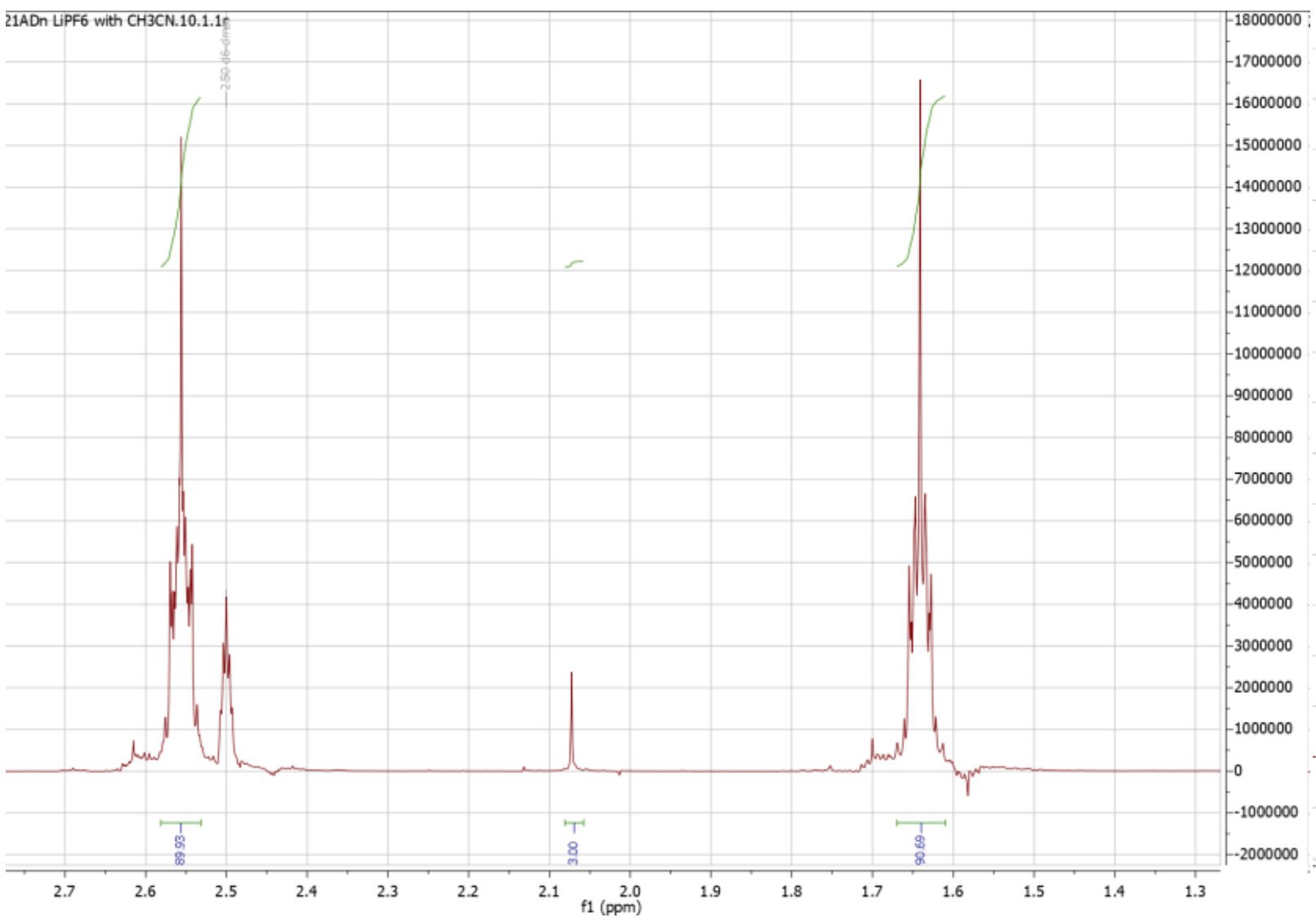

**(B)**

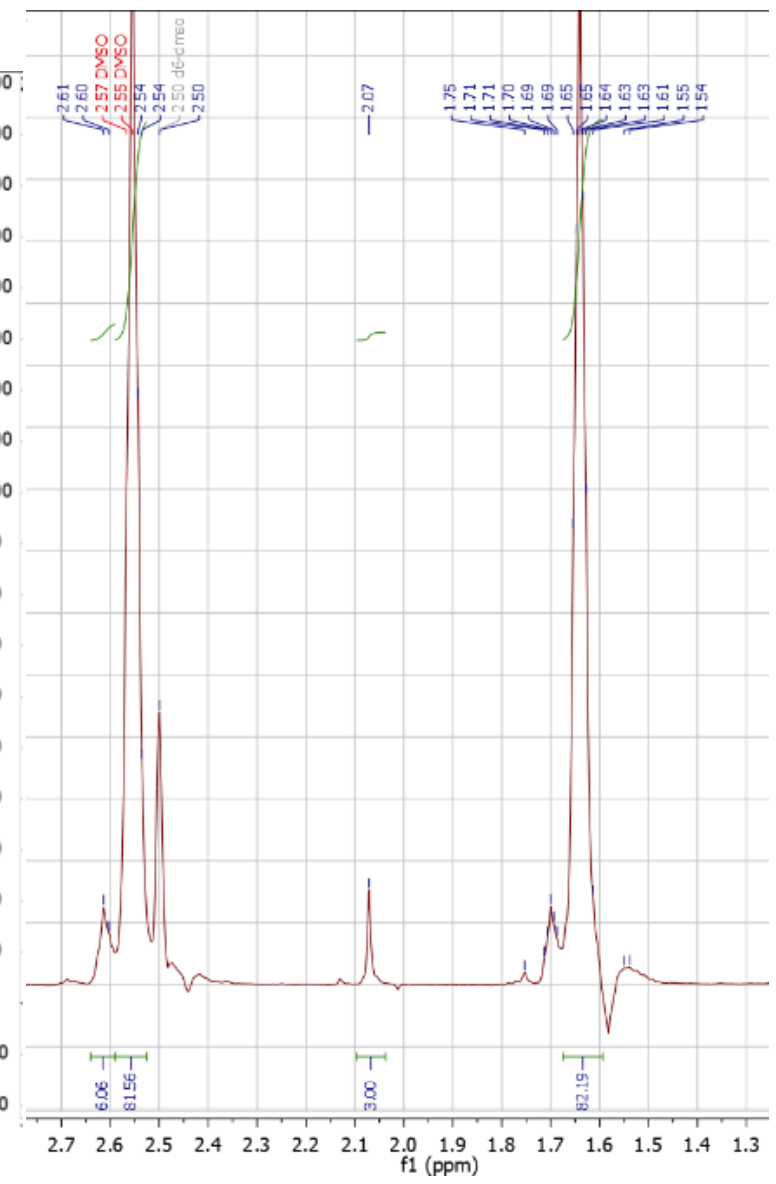

**(C)**

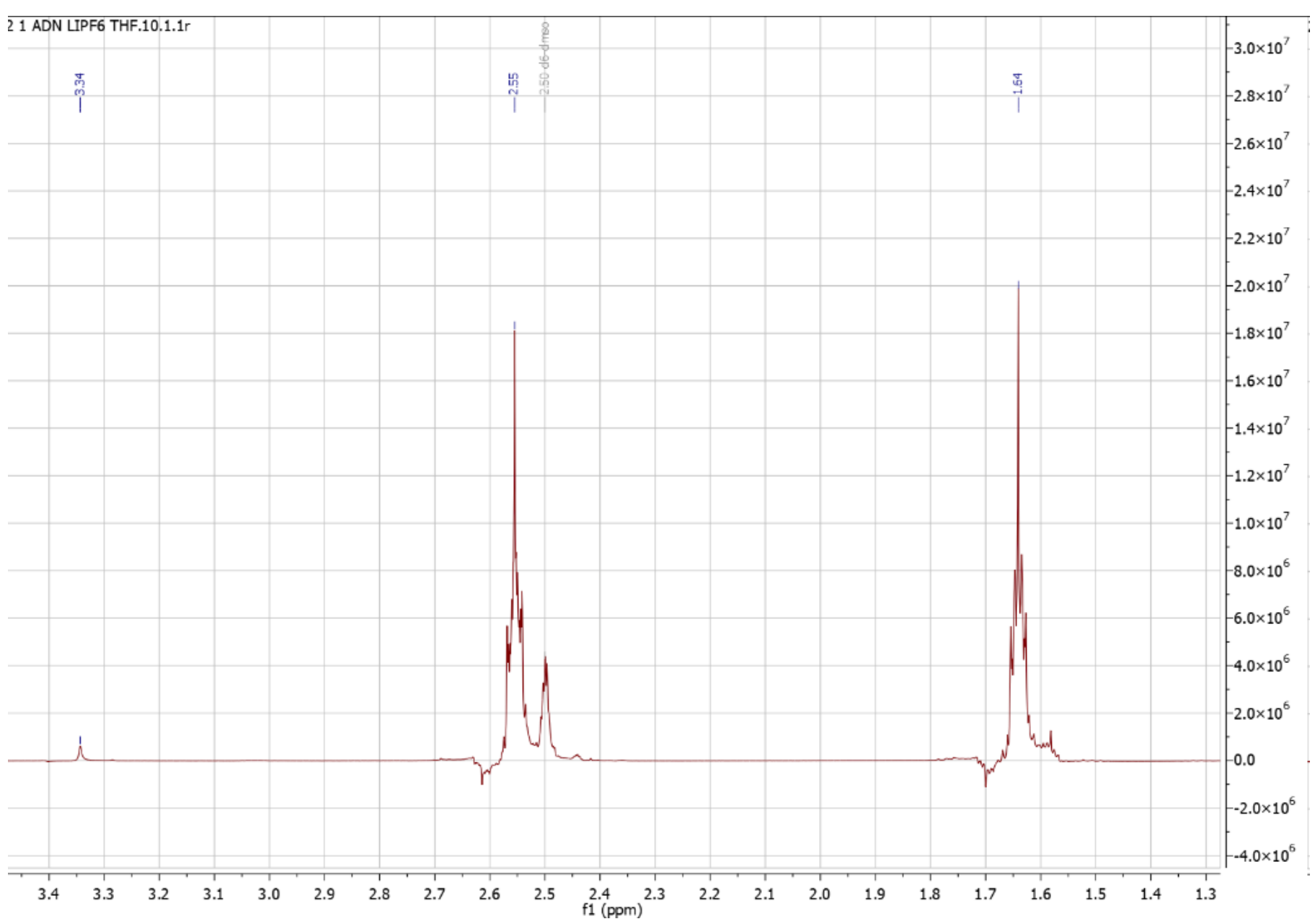

**Figure S5**. Proton NMR spectra of (**A**) large 2/1 crystal crystallized from AN (molar ratio of 8.6/1 Adpn/AN); (**B**) small 2/1 crystal crystallized from AN (molar ratio of 8.1/1 Adpn/AN); (**C**) small 2/1 crystal crystallized from THF. The molar ratio of Adpn/AN was 8.6/1 for the large and 8.1/1 for the small crystals, quantified by comparison of the methyl groups of AN and the $CH_2$ groups of Adpn.

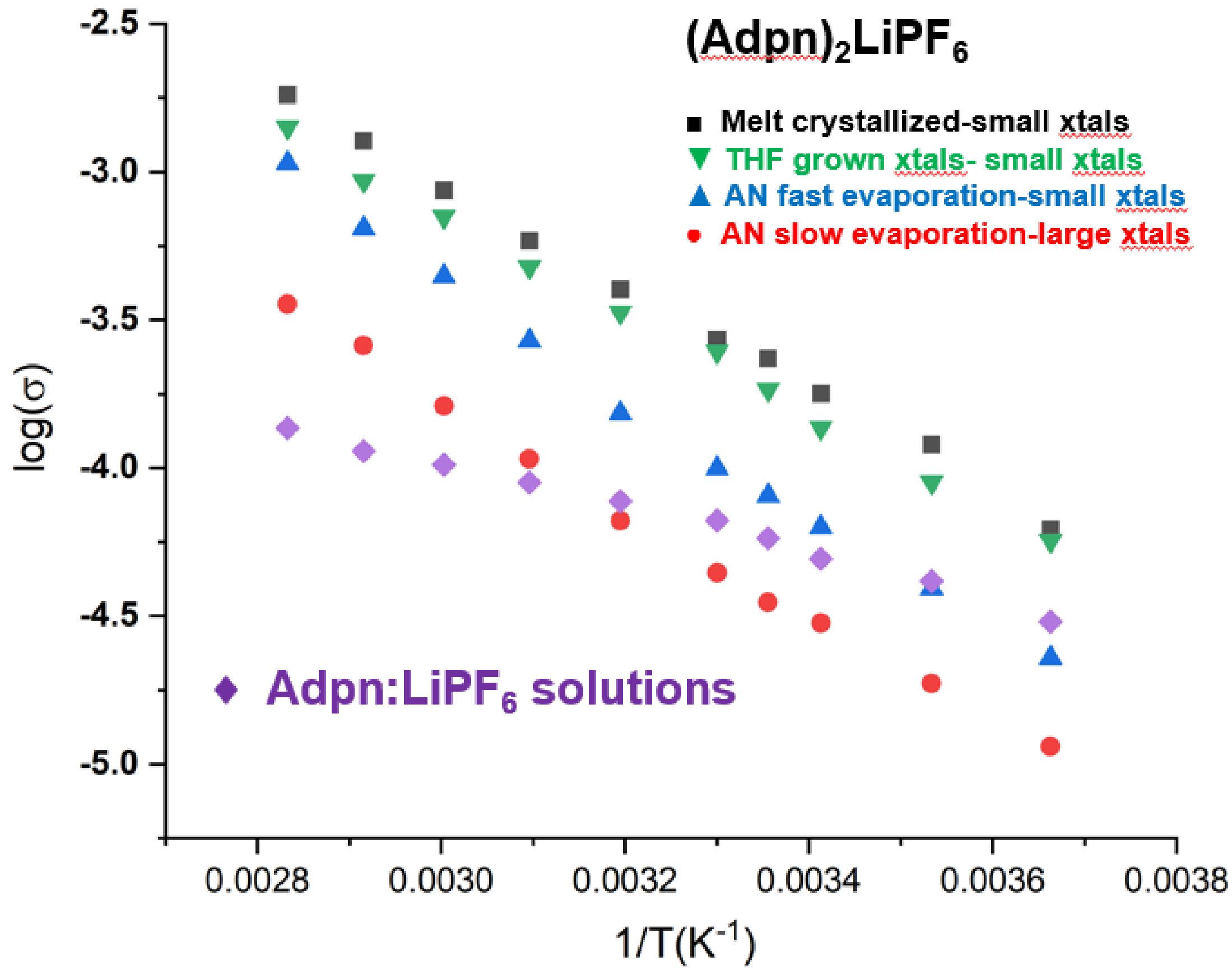

**Figure S6A**. Conductivity data, cooling cycle for $(Adpn)_2LiPF_6$ melt and solution (from AN) crystallized, and a saturated solution of $LiPF_6$ in Adpn.

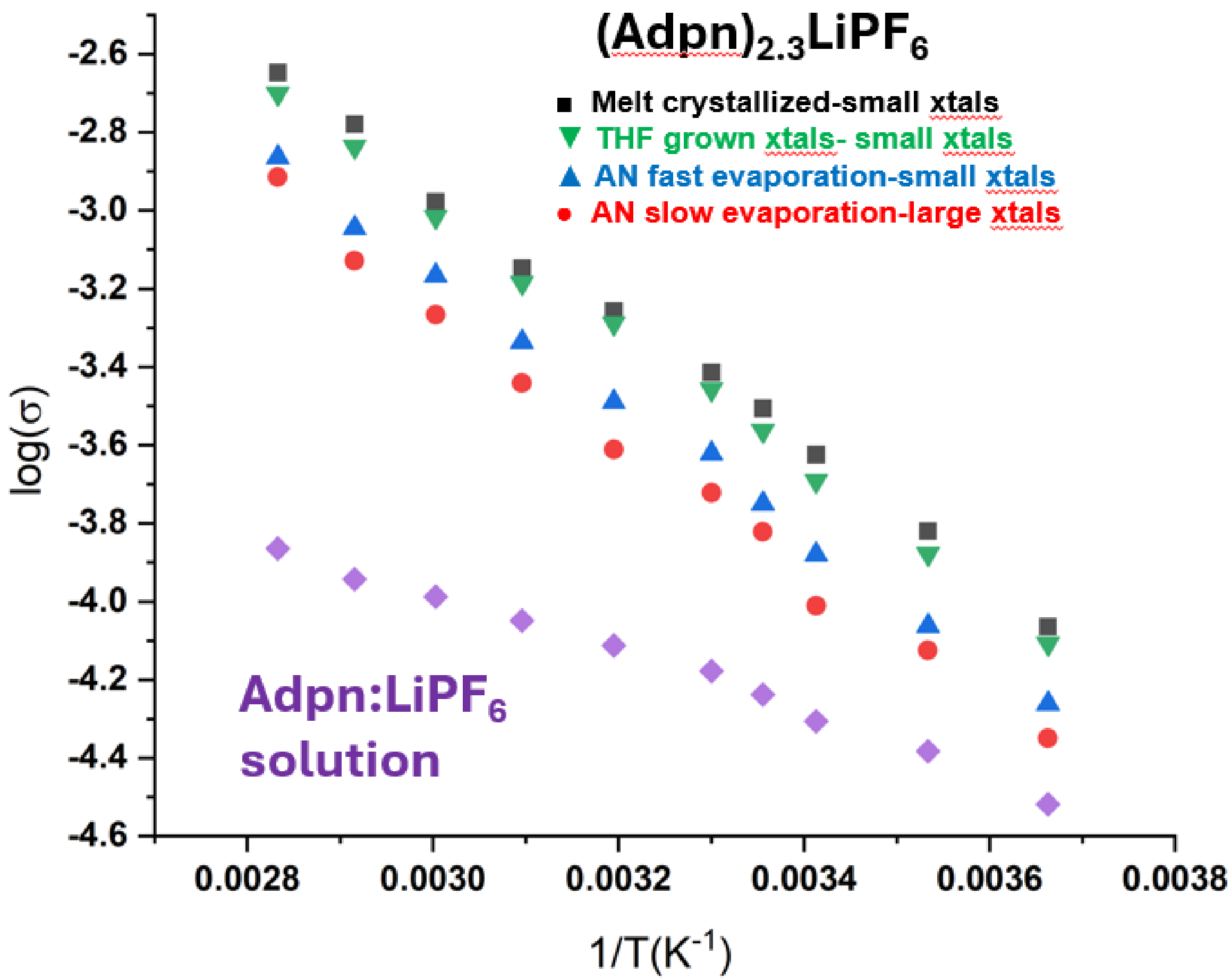

**Figure S6B**. Conductivity data, cooling cycle, for $(Adpn)_{2.3}LiPF_6$ melt and solution (from AN) crystallized, and a saturated solution of $LiPF_6$ in Adpn.

**Table S3.** Activation energy barriers from Arrhenius fit for temperature-dependent ionic conductivity data for heating cycles (shown in **Figure 2D,E**) for 2:1, 2.3:1 and saturated solution of Adpn:$LiPF_6$.

| **Sample type** | **$E_a$ ($kJ\text{-}mol^{-1}$)** | |
|---|---|---|
| | **2:1** | **2.3:1** |
| **Melt cystallized-small xtals** | 33.1 | 33.0 |
| **AN slow evapotation-large xtals** | 34.7 | 33.2 |
| **AN fast evapotation-small xtals** | 35.3 | 32.5 |
| **THF grown-small xtals** | 32.9 | 32.9 |
| **Saturated solution** | 14.9 | 14.9 |

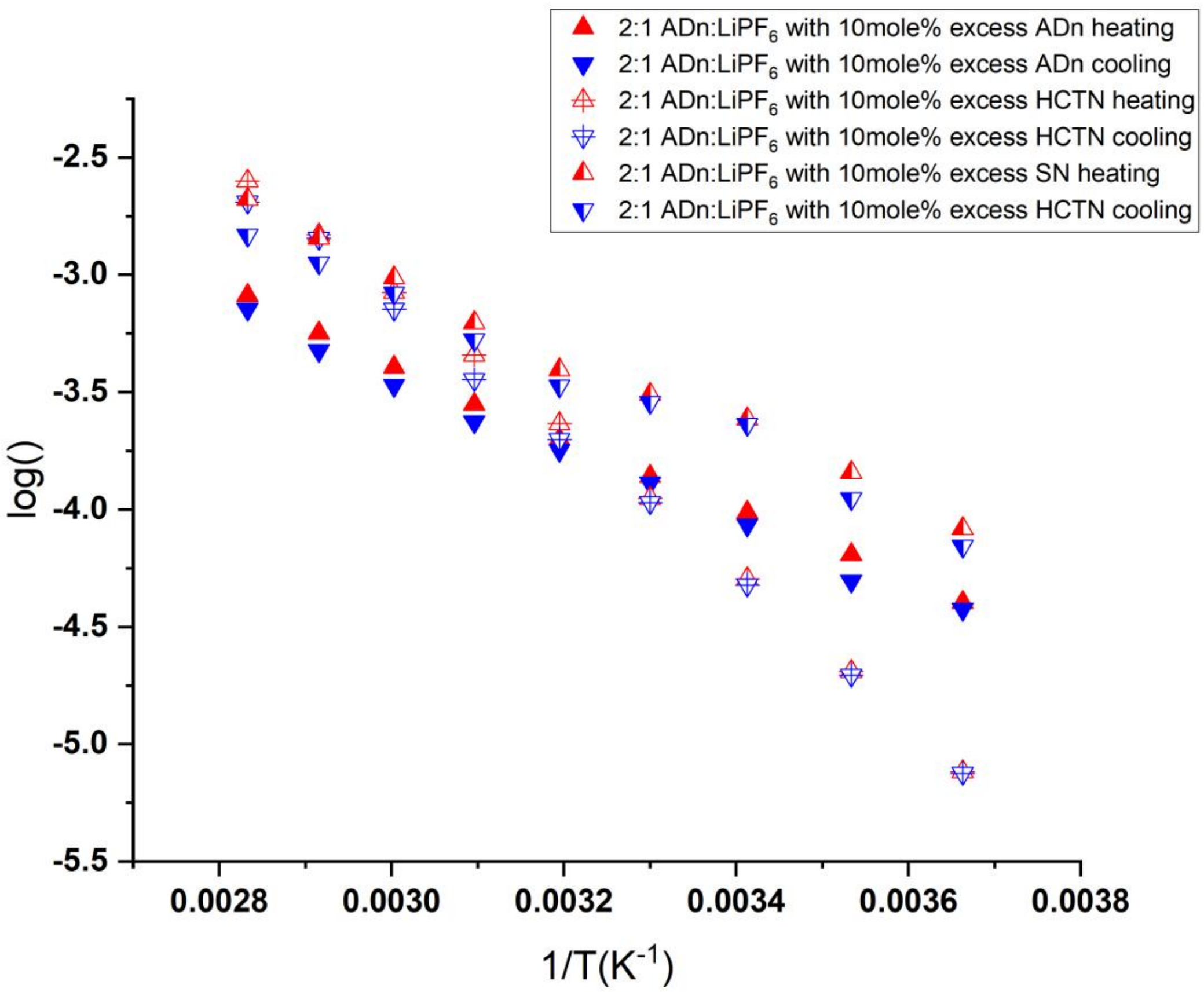

**Figure S7A**. Ionic conductivity, heating and cooling cycles, for Adpn:$LiPF_6$ with 10% excess Adiponitrile (Adpn), succinonitrile (SN) and 1, 3, 6 hexanetricarbonitrile (HTCN).

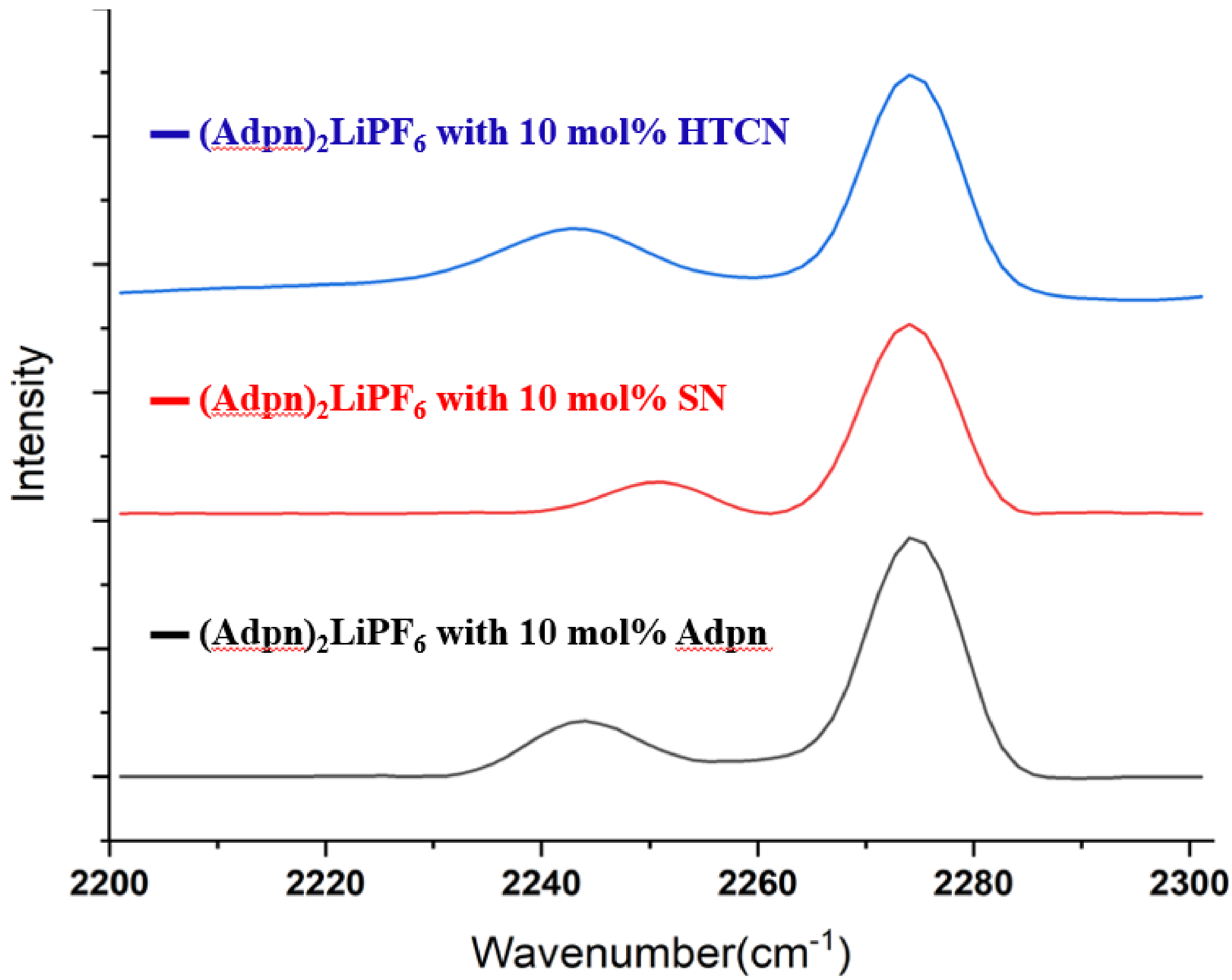

**Figure S7B**. Raman data of $(Adpn)_2LiPF_6$ crystals with 10 mol% excess Adipontrile (Adpn), succinonitrile (SN) and 1, 3, 6 hexanetricarbonitrile (HTCN).

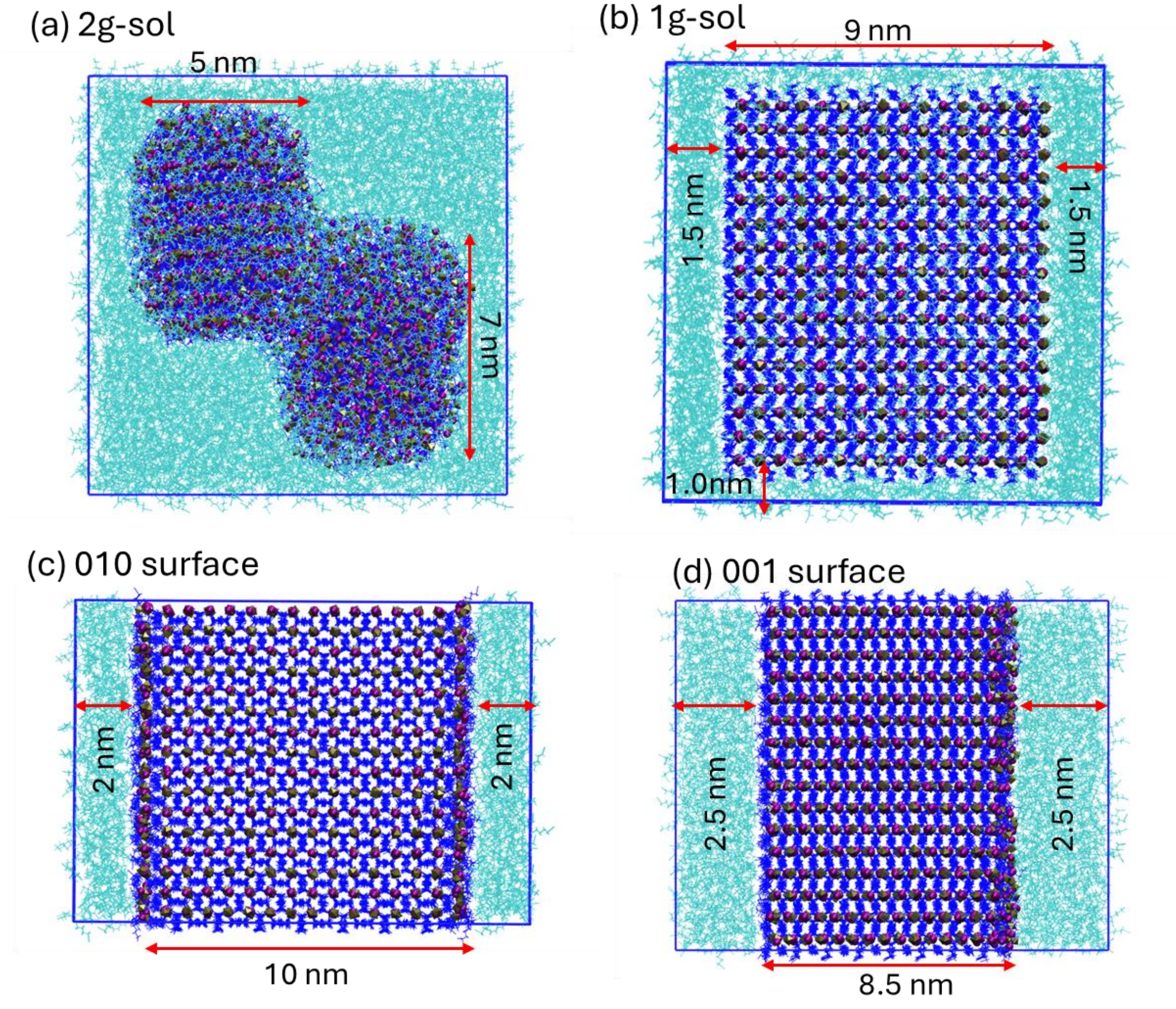

**Figure S8.** Four different excess-solvent models of $(Adpn)_{2+x}LiPF_6$ where x is the additional stoichiometry of excess Adpn: (a) Two small-grains (2g model), each with 500 formula units of $(Adpn)_2LiPF_6$ and 6286 excess Adpn molecules; (b) One-large grain (1g model) with 2048 formula units of $(Adpn)_2LiPF_6$ and 3080 excess Adpn molecules surrounding all sides; (c) 010 surface of crystals (in b crystallographic direction) with 2048 formula units solvated with 1680 excess Adpn molecules; (d) 001 surface of crystals (c crystallographic direction) with 2048 formula units of crystals solvated with 3360 excess Adpn molecules, where the left-side of the surface ions are completely coordinated with Adpn, and ions on the right side are half-coordinated. Colors: Purple sphere (●) Li; black octahedra $PF_6$ (*); blue lines ▬ Adpn (part of crystals); cyan lines ▬ excess Adpn.

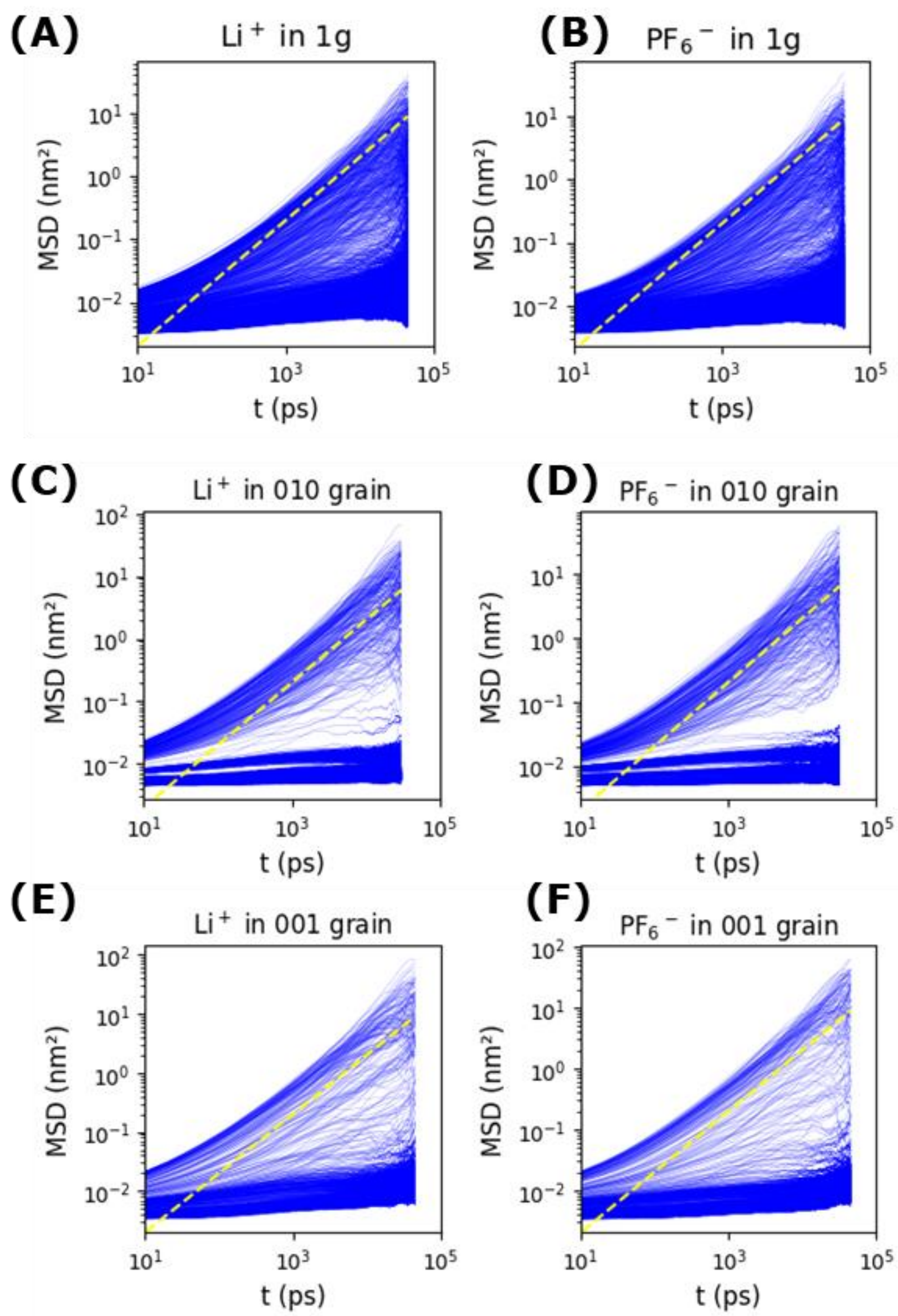

**Figure S9.** MSD *vs.* time plots of (**A**) $Li^+$ ions in 1g, (**B**) $PF_6^-$ ions in 1g, (**C**) $Li^+$ ions in 010 grain, (**D**) $PF_6^-$ ions in 010 grain, (**E**) $Li^+$ ions in 001 grain, (**F**) $PF_6^-$ ions in 001 grain. The yellow dotted line shows the slope of 1. Diffusion coefficients are only calculated for those "well-diffusive" carriers which have slopes of 1 (i.e., parallel to yellow line).